\documentclass{jfm}
\usepackage{graphicx}
\usepackage{epstopdf, epsfig}
\usepackage{xcolor}
\usepackage{mathtools}

\newcommand{\mylab}[3]{\raisebox{#2}[0mm][0mm]{\makebox[0mm][l]{\hspace*{#1}#3}}}

\definecolor{matlabmagenta}{rgb}{1.0, 0.0, 1.0}

\graphicspath{{Figures_for_submission/}}

\shorttitle{Virtual-origin framework for turbulence over drag-altering surfaces}
\shortauthor{J. I. Ibrahim, G. G{\'o}mez-de-Segura, D. Chung and R. Garc{\'i}a-Mayoral}

\title{The smooth-wall-like behaviour of turbulence over drag-altering surfaces: a~unifying virtual-origin framework}

\author{Joseph I. Ibrahim\aff{1}, Garazi G{\'o}mez-de-Segura\aff{1}, Daniel Chung\aff{2} \and Ricardo Garc{\'i}a-Mayoral\aff{1}\corresp{\email{r.gmayoral@eng.cam.ac.uk}}}

\affiliation{\aff{1}Department of Engineering, University of Cambridge, Trumpington St, Cambridge CB2 1PZ, UK\\
\aff{2}Department of Mechanical Engineering, University of Melbourne, Victoria 3010, Australia}

\begin{document}

\maketitle

\begin{abstract}
We examine the effect on near-wall turbulence of displacing the apparent, virtual origins perceived by different components of the overlying flow. This mechanism is commonly reported for drag-altering textured surfaces of small size. For the particular case of riblets, \citeauthor{Luchini1991} (\textit{J. Fluid Mech.}, vol.\ 228, 1991, pp.\ 87--109) proposed that their effect on the overlying flow could be reduced to an offset between the origins perceived by the streamwise and spanwise velocities, with the latter being the origin perceived by turbulence. Later results, particularly in the context of superhydrophobic surfaces, suggest that this effect is not determined by the apparent origins of the tangential velocities alone, but also by the one for the wall-normal velocity. To investigate this, the present paper focuses on direct simulations of turbulent channels imposing different virtual origins for all three velocity components using Robin, slip-like boundary conditions, {and also using} opposition control. Our simulation results support that the relevant parameter is the offset between the virtual origins perceived by the mean flow and turbulence. {When using Robin, slip-like boundary conditions, the virtual origin for the mean flow is determined by the streamwise slip length. Meanwhile, the virtual origin for turbulence results from the combined effect of the wall-normal and spanwise slip lengths.} The slip experienced by the streamwise velocity fluctuations, in turn, has a negligible effect on the virtual origin for turbulence, and hence the drag, at least in the regime of drag reduction. This suggests that the origin perceived by the quasi-streamwise vortices, which induce the cross-flow velocities at the surface, is key in determining the virtual origin for turbulence, while that perceived by the near-wall streaks, which are associated with the streamwise velocity fluctuations, plays a secondary role. {In this framework, the changes in turbulent quantities typically reported in the flow-control literature are shown to be merely a result of the choice of origin, and are absent when using as origin the one experienced by turbulence. Other than this shift in origin, we demonstrate that turbulence thus remains essentially smooth-wall-like. A simple expression can predict the virtual origin for turbulence in this regime.} The effect can {also} be reproduced a priori by introducing the virtual origins into a smooth-wall eddy-viscosity {framework.}
\end{abstract}


\section{Introduction}\label{sec:introduction}

Some textured surfaces, such as riblets~\citep{Walsh1984}, superhydrophobic surfaces~\citep{Rothstein2010} and anisotropic permeable substrates~\citep{Gomez-de-Segura2019}, are designed to manipulate the flow to modify the turbulent skin-friction drag compared to a smooth surface. Like many turbulent flow-control techniques, including active ones, these surfaces typically aim to manipulate the near-wall cycle~\citep{Hamilton1995,Waleffe1997}  due to its key role in the generation of turbulent skin friction~\citep{Jimenez1999}. For example, for drag-reducing surfaces of small texture size, the general idea is to impede the flow in the streamwise direction less than the cross flow. The net effect can then be thought of as a  relative outward displacement of the quasi-streamwise vortices  with respect to the mean flow~\citep{Jimenez1994,Luchini1996}. This reduces the local turbulent mixing of streamwise momentum and, therefore, the skin-friction drag~\citep{Orlandi1994}. 

{Provided that the direct effect of the texture is confined to the near-wall region, the classical theory of wall turbulence postulates that the change in drag is manifested as a constant shift in the mean velocity profile, $\Delta U^+$, experienced by the flow above the near-wall region~\citep{Clauser1956,Spalart2011,Garcia-Mayoral2019}.} In this paper, we choose $\Delta U^+ > 0$ to denote drag reduction. However, we note that in the roughness community, {the sign is typically reversed and $\Delta U^+$, referred to as the roughness function, is positive when drag increases~\citep{Jimenez2004}}. {The} superscript `$+$' indicates scaling in wall units, i.e. normalisation by the friction velocity $u_\tau = \sqrt{\tau_w/\rho}$  and the kinematic viscosity $\nu$, where $\tau_w$ is the wall shear stress and $\rho$ is the density.  \citet{Jimenez1994} and \citet{Luchini1996} proposed that $\Delta U^+$ could depend only on the height difference between two apparent virtual origins imposed by the surface on the flow. In their original framework, these would be the virtual origins perceived by the streamwise and spanwise velocity, respectively. However, the results of a recent preliminary study by  \citet{Gomez-de-Segura2018a} suggest that, in general, to fully describe the effects of a textured surface on the flow, it is also necessary to consider an apparent virtual origin for the wall-normal velocity. The important role of the wall-normal velocity has also been observed in turbulent flows over rough surfaces, for which $\Delta U^+$ shows correlation with the wall-normal Reynolds stress at the roughness crests~\citep{Orlandi2008,Orlandi2013,Orlandi2019}.

{The aim of the present work is to develop a unifying virtual-origin framework in which the effect of the surface texture or flow-control strategy can be reduced to a relative offset between the virtual origins perceived by different components of the flow, and which components those would be.  This effect has been observed in direct numerical simulations (DNSs) of certain textured surfaces~\citep{Gomez-de-Segura2018c}, and also in DNSs with active opposition control~\citep{Choi1994}.} {{We impose such origins using Robin, slip-length-like boundary conditions. This has been} proposed by \citet{Gomez-de-Segura2020} as {a} simple and effective method. We note that, formally, first-order homogenisation produces Robin boundary conditions for the tangential velocities alone, and a non-zero transpiration arises only for second- or higher-order expansions. Our boundary conditions should thus not be viewed as equivalent boundary conditions in the sense of homogenisation, but purely as a method to impose virtual origins.} {Although not the focus of this paper, we refer the reader to the works of \citet{Bottaro2019}, \citet{Bottaro2020} and \citet{Lacis2020} for the discussion on how to obtain such equivalent boundary conditions for actual textures. The expansion in homogenisation is typically done for the small parameter given by the ratio of the texture size to the flow thickness. One difficulty in turbulent flows, however, is that the ratio would need to remain small even for the smallest length scales in the flow. These would typically be the diameter of the near-wall quasi-streamwise vortices or their height above the surface, both of order 15 wall units~\citep{Robinson1991,Schoppa2002}. This implies that for the expansion to converge the texture size would need to be even smaller. Most textures in that size range, however, behave as hydraulically smooth~\citep{Jimenez1994}.} {Alternatively, the} focus of this work is the extent to which the {velocities perceiving different} virtual origins modify the dynamics {of turbulence}. We also {aim} to determine if it is possible to predict the shift in the mean velocity profile, $\Delta U^+$, a priori from the apparent virtual origins perceived by the flow. {While the observation that some textures produce such an effect is the motivation behind our work, it is beyond the scope of the present paper to quantify this effect for specific textures, although some preliminary work on this can be found in \citet{Gomez-de-Segura2018c}. It is also beyond our scope to derive equivalent boundary conditions for specific textures, or to establish the connection between such equivalent conditions and the observed virtual-origin effect.}


The paper is organised as follows. In \S\,\ref{sec:theory}, we present and discuss the current understanding of how small-textured surfaces modify the drag compared to a smooth surface by imposing apparent virtual origins to the flow velocity components. Then, \S\,\ref{sec:numerics} details the numerical method of our DNSs and summarises the series of simulations we conduct. The results are presented in \S\,\ref{sec:results}, where we discuss in detail the effect on the flow of imposing different virtual origins for all three velocity components. We also propose, from physical and empirical arguments, an expression that can be used to predict $\Delta U^+$ from the apparent virtual origins a priori. Section~\ref{sec:opposition} {discusses our} results on opposition control~\citep{Choi1994}, which suggest that certain active flow-control techniques can also be interpreted in terms of virtual origins. In \S\,\ref{sec:eddyVisc}, we present a theoretical framework that reproduces a priori the effect observed in our simulations. We summarise our key findings in the final section.

\section{Mean-velocity shift, drag, and virtual origins}\label{sec:theory}

When a surface produces the aforementioned shift in the mean velocity profile, $\Delta U^+$, in the log and outer regions of the flow, we would have~\citep{Clauser1956}
\begin{equation}
U^+ = \frac{1}{\kappa}\log y^+ + B + \Delta U^+,\label{eq:loglaw}
\end{equation}
where $U$ is the mean streamwise velocity and $y$ is the distance from the wall. {The von K{\'a}rm{\'a}n constant, $\kappa$,  remains unchanged, and so does the function $B$, which contains both the $y$-intercept of the log law and the wake function.} If the texture size remains constant in wall units and the effect of the texture is confined to the near-wall region, the consensus is that $\Delta U^+$ is essentially independent of the friction Reynolds number, as discussed by \citet{Garcia-Mayoral2011a} and \citet{Spalart2011} in the context of riblets. The shift $\Delta U^+$ produced by some other flow-control strategies, such as spanwise wall oscillation, has also been reported to be essentially independent of the Reynolds number, so long as the parameters that describe the control remain constant in wall units~\citep{Gatti2016}. However, regardless of the control strategy, $\Delta U^+$ is weakly affected by the modulation of the local viscous length scale by the intensity of the large scales in the flow, an effect that becomes more significant at larger Reynolds numbers~\citep{Mathis2009,Zhang2016}. This effect is typically of the order of a few per cent at the Reynolds numbers of engineering applications~\citep{Hutchins2015,Chernyshenko2019}, so we will neglect it here.

In turn, the change in drag is strictly dependent on the Reynolds number. The skin friction coefficient is defined as $c_f = 2\tau_w/(\rho U_\delta^2)=2/U_\delta^{+2}$, where $U_\delta$ is the reference velocity. The choice of $U_\delta$ depends on the type of flow considered. Typically for external flows, $U_\delta$ would be the free-stream velocity, while for internal flows it would be the bulk velocity. For internal flows, $U_\delta$ can also be the centreline velocity, for comparison with external flows of interest~\citep{Garcia-Mayoral2011b}. Using the subscript `0' to denote reference smooth-wall values, the drag reduction, $\mbox{\textit{DR}}$, can then be expressed as the relative decrease in $c_f$ compared to that for a smooth wall, $c_{f_0}$,
\begin{equation}
\mbox{\textit{DR}} = -\frac{\Delta c_f}{c_{f_0}},\label{eq:DR1}
\end{equation} 
where $\Delta c_f = c_f - c_{f_0}$. As discussed by \citet{Garcia-Mayoral2019}, care must be taken when quoting values of drag reduction achieved by textured surfaces. For example, the corresponding position of the reference smooth wall, particularly in the case of internal flows, can imply a potentially significant change in the hydraulic radius between experiments and applications, resulting in values of $\mbox{\textit{DR}}$ not directly attributable to the texture. 

From (\ref{eq:DR1}) and the definition of $c_f$ above, $\mbox{\textit{DR}}$ can be given in terms of $\Delta U^+$ as follows. Provided the Reynolds number is large enough that {outer-layer similarity is} observed and (\ref{eq:loglaw}) holds, it follows from (\ref{eq:loglaw}) that $U^+_\delta = U^+_{\delta_0} + \Delta U^+$. Then (\ref{eq:DR1}) can be written as~\citep{Garcia-Mayoral2019}
\begin{equation}
\mbox{\textit{DR}} = 1 - \left(\frac{1}{1+\Delta U^+/U^+_{\delta_0}} \right)^2. \label{eq:DR2}
\end{equation} 
Since $U^+_{\delta_0}$ depends on the Reynolds number, so too will the drag, for $\Delta U^+$ fixed. {This leaves $\Delta U^+$ as the Reynolds-number independent means of quantifying the change in drag and extrapolating laboratory results to applications~\citep{Spalart2011,Garcia-Mayoral2019}.}

\begin{figure}
\centering    
\includegraphics[width=0.7\textwidth]{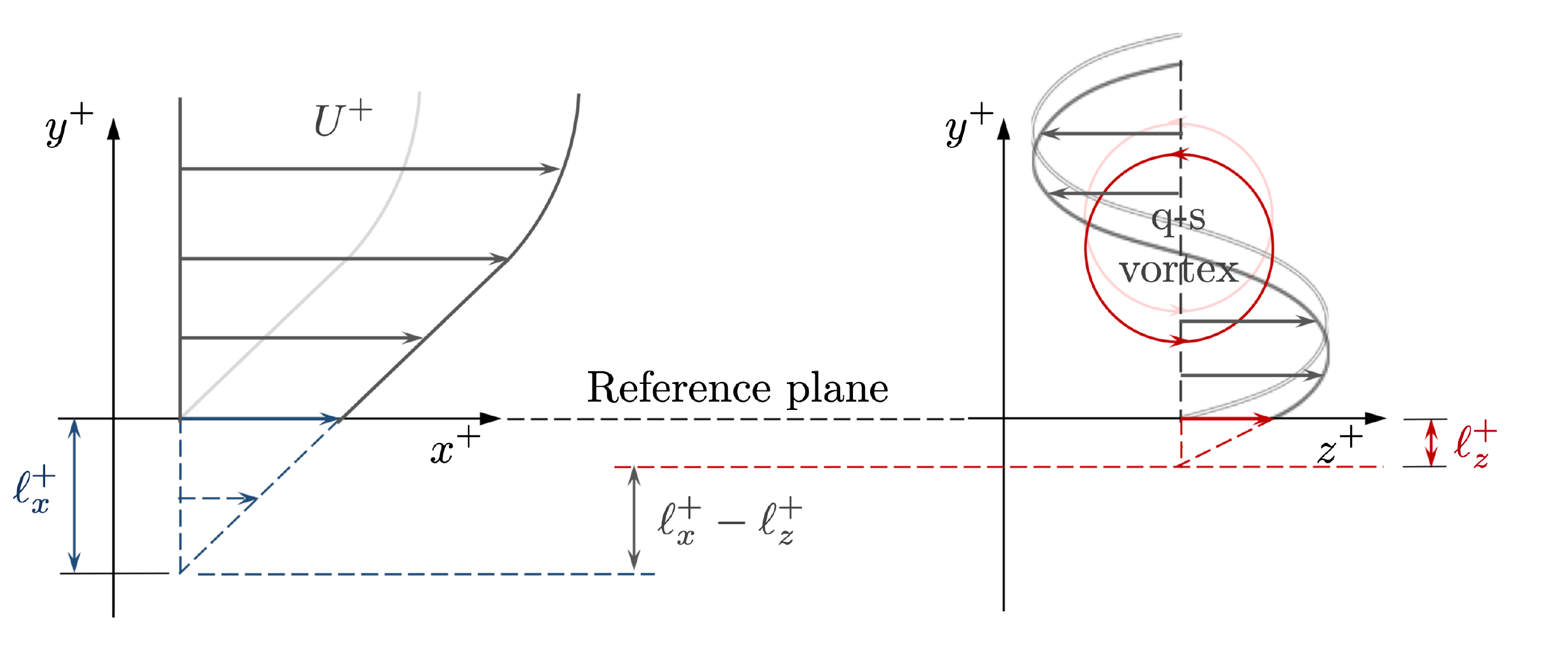}
\mylab{-9.6cm}{3.6cm}{($a$)}%
\mylab{-4.1cm}{3.6cm}{($b$)}%
\caption{Schematic of ($a$) streamwise and ($b$) spanwise slip lengths,  $\ell_x^+$ and $\ell_z^+$, and the corresponding virtual origins at $y^+=-\ell_x^+$ and $y^+=-\ell_z^+$. A quasi-streamwise vortex (q-s vortex), inducing a spanwise velocity $w^+$, is sketched in ($b$). Grey profiles indicate smooth-wall behaviour with the wall located at the reference plane.}
\label{fig:VOs}
\end{figure}

{We now  discuss} the way in which surfaces with small texture produce a shift in the mean velocity profile, $\Delta U^+$, and hence modify the drag. The early studies focused on the drag reduction mechanism in riblets, {but the analysis can easily be extended to other surfaces.} {We use} $x$, $y$ and $z$ as the streamwise, wall-normal and spanwise coordinates, respectively, and $u$, $v$ and $w$ as their corresponding velocity components. {We also use} $\delta$ to refer to the flow thickness, which, depending on the application, would correspond to the channel half-height, boundary layer thickness or pipe radius. \citet{Bechert1989} originally suggested that, for riblets, the streamwise velocity experiences an apparent, no-slip wall, or virtual origin, at a depth $\ell_x$  below the riblet tips, which they called the `protrusion height'. This concept is depicted in figure~\ref{fig:VOs}($a$). Note that, in the superhydrophobic-surface community, $\ell_x$ is often referred to as the streamwise slip length, and, in this paper, we will use the term `slip length' instead of `protrusion height'. {Defining for convenience} the reference plane $y=0$ to be located at the riblet tips, and noting that the velocity profile is essentially linear in the viscous sublayer, this is equivalent to a Navier slip condition {at $y=0$} of the form
\begin{equation}
u = \ell_x\frac{\partial u}{\partial y}. \label{eq:navierSlip}
\end{equation}
The virtual origin for the streamwise velocity is then at $y = - \ell_x$. The streamwise flow thus perceives an apparent, no-slip wall at a distance $\ell_x$ below the riblet tips, $y=0$. In wall units, the mean streamwise shear $\mathrm{d} U^+/\mathrm{d} y^+ = 1$ very near the wall, and so (\ref{eq:navierSlip}) becomes $U^+(y^+=0) = \ell_x^+$. In other words, the slip velocity experienced by the mean flow is equal to the streamwise slip length expressed in wall units, so the concept of the slip length $\ell_x^+$ is often used interchangeably with that of the slip velocity $U_s^+(y^+=0) = \ell_x^+$.

The {above} implies that the spanwise velocity, generated by the quasi-streamwise vortices of the near-wall cycle, would perceive an origin at the riblet tips $y=0$. In general, this is not the case, and the vortices would instead perceive an origin at some distance below the plane $y=0$. \citet{Luchini1991} proposed, therefore, that it would also be necessary to consider a spanwise slip length, $\ell_z$, to properly describe the effect of riblets on the flow with respect to the reference plane $y=0$. Again, this would be equivalent to a Navier slip condition {at $y=0$} on the spanwise velocity,
\begin{equation}
w=\ell_z\frac{\partial w}{ \partial y},
\end{equation}
where $y = -\ell_z$ would be the location of the virtual origin for the spanwise turbulent fluctuations generated primarily by the quasi-streamwise vortices, as portrayed in figure~\ref{fig:VOs}($b$). \citet{Luchini1991} {concluded that} the only important parameter in determining the drag reduction due to riblets would be the difference between these two virtual origins, i.e. $\ell_x - \ell_z$. The physical justification for this is that the origin perceived by the quasi-streamwise vortices would likely set the origin perceived by the whole turbulence dynamics, as proposed by \citet{Luchini1996}. In other words, the dynamics of turbulence would be displaced `rigidly' with the vortices and they would both perceive a virtual origin at the same depth, which would be at $y=-\ell_z$ in the above framework. {Note} that even though this analysis was conducted in the context of riblets, it is also valid for any small-textured surface that generates different virtual origins for the streamwise and spanwise velocities. The relationship between $\Delta U^+$ and $\ell_x - \ell_z$ was studied further by \citet{Luchini1996}, for riblets, and by \citet{Jimenez1994}, in a texture-independent framework, and they concluded that $\Delta U^+ \propto \ell_x^+ - \ell_z^+$ for $\ell^+_x,\ell^+_z \lesssim 1$, with a constant of proportionality of order 1. \citet{Garcia-Mayoral2019} argued recently that the constant of proportionality must necessarily be 1, i.e.\ $\Delta U^+ = \ell_x^+ - \ell_z^+$. In practice, the requirement $\ell^+_x,\ell^+_z \lesssim 1$ can be somewhat relaxed, provided that the overlying flow perceives only the homogenised effect of the texture. This would require the texture to be smaller than the overlying turbulent eddies in the flow.

\begin{figure}
\centering    
\includegraphics[width=0.48\textwidth]{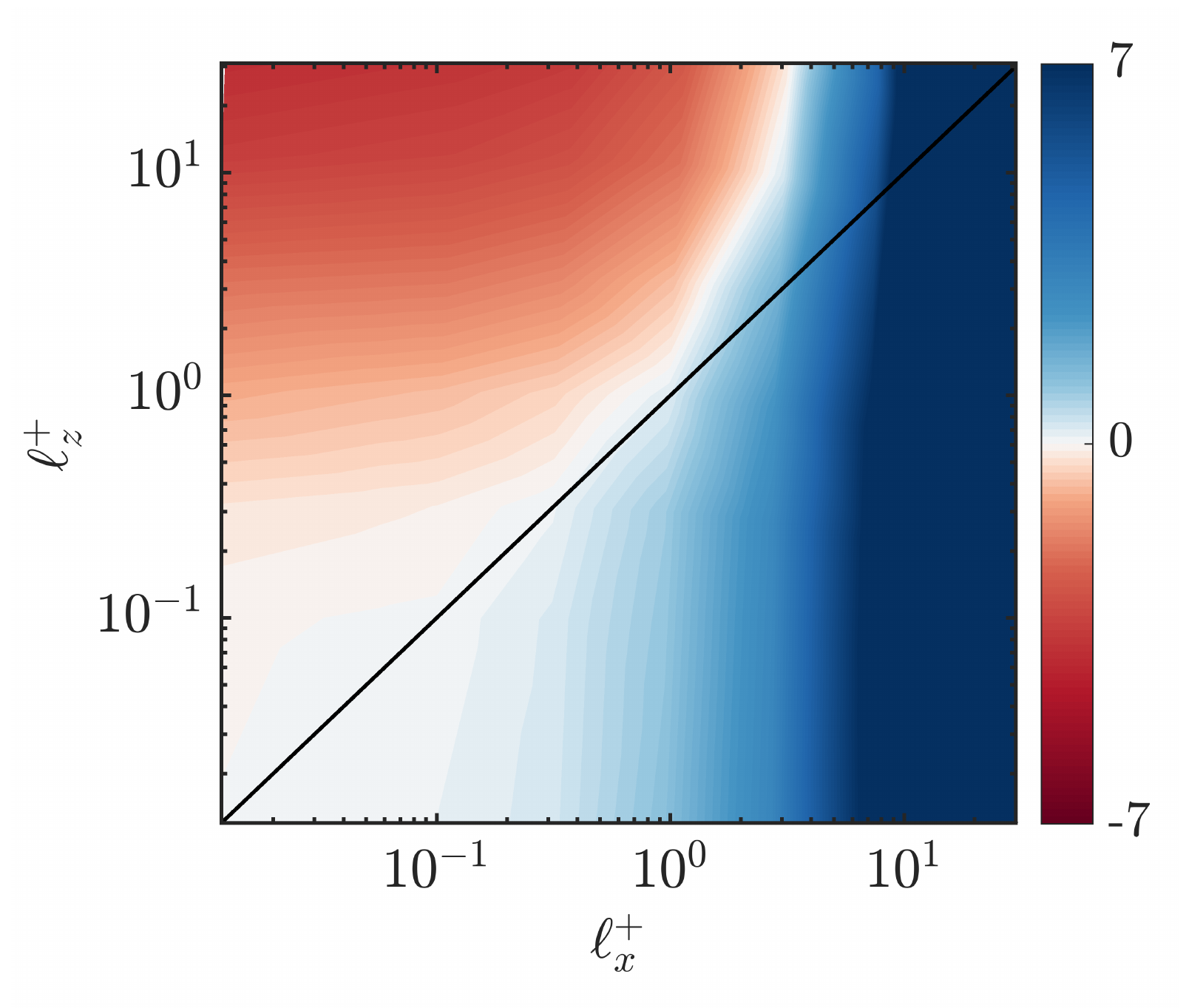}\hspace{6pt}
\includegraphics[width=0.4\textwidth]{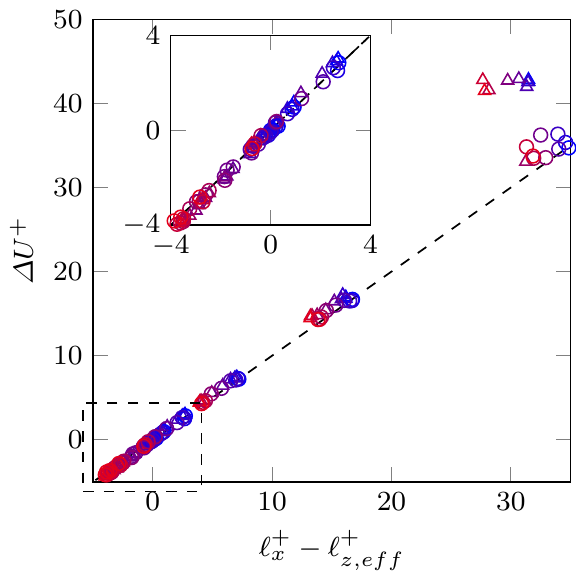}
\mylab{-12.1cm}{5.3cm}{($a$)}%
\mylab{-5.5cm}{5.3cm}{($b$)}%
\caption{(\textit{a}) Map of $\Delta U^+$ for different slip lengths, $\ell_x^+$ and $\ell_z^+$, from~\citet{Busse2012} starting from a smooth-wall friction Reynolds number $\Rey_{\tau,0}  = 180$. Black solid line, $\ell_x^+=\ell_z^+$. (\textit{b}) $\Delta U^+$ as a function of $\ell_x^+ - \ell_{z,eff}^+$, using the same data as in (\textit{a}). Triangles, simulations at $\Rey_{\tau,0} = 180$; circles, simulations at $\Rey_{\tau,0}  = 360$. From blue to red, the spanwise slip length increases. Dashed line, $\Delta U^+ = \ell_x^+ - \ell_{z,eff}^+$.  Adapted from \citet{Fairhall2018}.} 
\label{fig:BS_saturation}
\end{figure}

However, when the spanwise slip length generated by a surface becomes larger than a few wall units, the effect of $\ell_z^+$ on $\Delta U^+$ starts to saturate \citep{Min2004,Fukagata2006}. \citet{Busse2012} conducted a parametric study for a wide range of streamwise and spanwise slip lengths and showed that, for $\ell_x^+ \gtrsim 4$, drag is reduced for all values of $\ell_z^+$, as shown in figure~\ref{fig:BS_saturation}(\textit{a}). In this regime, $\Delta U^+$ is no longer simply proportional to the difference between the streamwise and spanwise slip lengths. If it were, the contours in figure~\ref{fig:BS_saturation}(\textit{a}) would be symmetric about the diagonal line. \citet{Fairhall2018} have since shown that this saturation can be accounted for with an `effective' spanwise slip length, $\ell^+_{z,eff}$, which is empirically observed to be
\begin{equation}
\ell^+_{z,eff} \approx \frac{\ell_z^+}{1+\ell_z^+/4}.\label{eq:lzeff}
\end{equation} 
The change in drag would then be $\Delta U^+ =  \ell_x^+ - \ell_{z,eff}^+$. For $\ell_z^+ \lesssim 1$, $\ell_{z,eff}^+ \approx \ell_z^+$, recovering the above expression that $\Delta U^+ = \ell_x^+ - \ell_z^+$, while for large values of $\ell_z^+$, $\ell_{z,eff}^+$ asymptotes to 4. {From  (\ref{eq:lzeff}), if the spanwise slip length was $\ell_z^+ = 1$, the effective spanwise slip length would be a similar $\ell_{z,eff}^+ = 0.8$.  However, $\ell_z^+ = 2$ would only yield $\ell_{z,eff}^+ = 1.3$, a reduction of more than 30\%. This suggests that the prediction for $\Delta U^+$ from slip conditions obtained from homogenisation begins to break down already for spanwise slip lengths as small as 2 wall units. This can typically lie in the hydraulically smooth regime, i.e.\ models that consider tangential slip alone can cease to hold before $\Delta U^+$ reaches values of relevance.} From (\ref{eq:lzeff}), the two-dimensional parametric space $(\ell_x^+,\ell_z^+)$ in figure~\ref{fig:BS_saturation}(\textit{a}) can be fitted to a single curve, as shown in figure~\ref{fig:BS_saturation}(\textit{b}). Note that this would extend the validity of (\ref{eq:lzeff}) from $\ell_x^+,\ell_z^+ \sim 1$ to $\ell_x^+,\ell_z^+ \sim 30$ at least, so long as the flow only perceived the underlying texture in a homogenised fashion. There is some deviation for the cases at the lower smooth-wall friction Reynolds number, $\Rey_{\tau,0} = 180$, when the streamwise slip length is large, $\ell_x^+ \sim 100$. This was likely a low-Reynolds-number effect associated with the flow relaminarising, since the simulations were conducted at constant mass flow rate, so that a large drag reduction resulted in a significant decrease in $\Rey_\tau$.

\begin{figure}
\centering    
\includegraphics[width=0.7\textwidth]{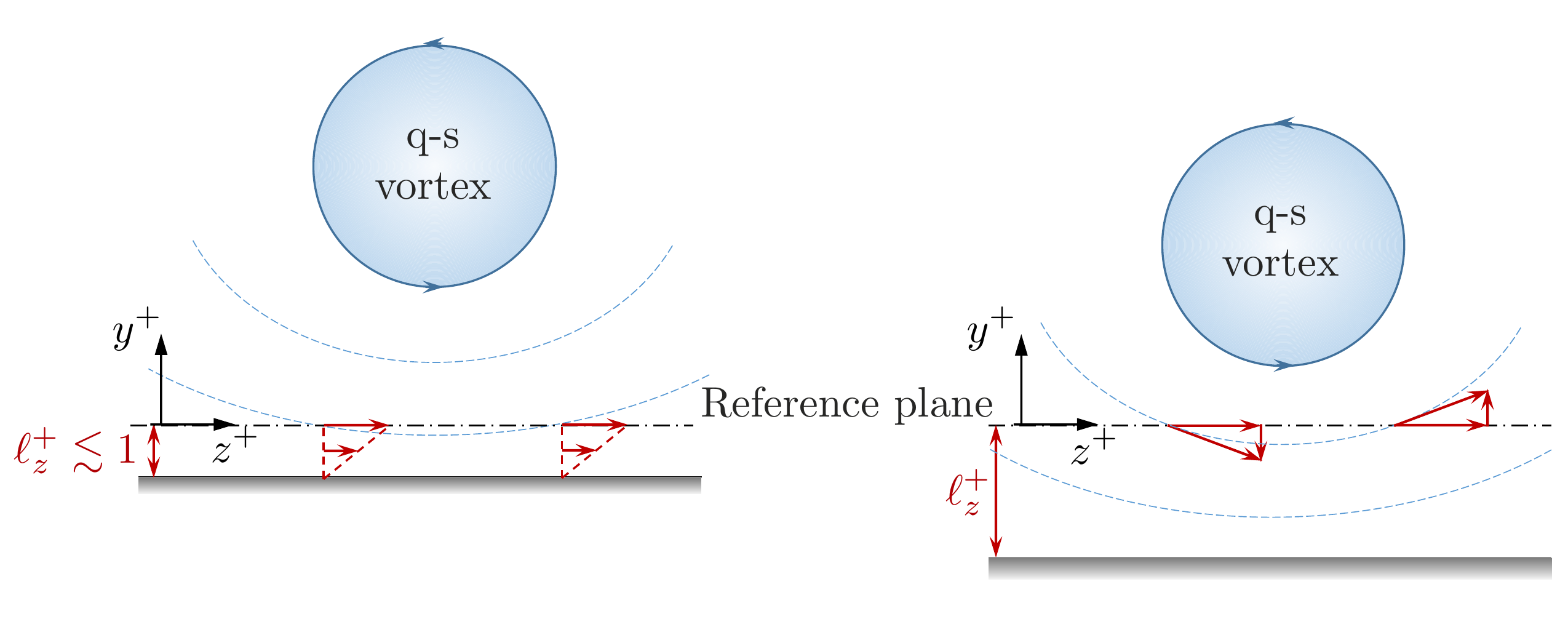}
\mylab{-9cm}{3.5cm}{($a$)}%
\mylab{-4cm}{3.5cm}{($b$)}%
\caption{Schematic of spanwise and wall-normal velocities induced by quasi-streamwise vortices (q-s vortex) at the reference plane for ($a$) virtual origins $\ell_z^+ \lesssim 1$ and ($b$) larger virtual origins. Shaded grey regions indicate the apparent smooth wall perceived by the vortex.} 
\label{fig:v_vortex}
\end{figure}

\citet{Gomez-de-Segura2018a} have recently investigated the cause for this saturation in the effect of $\ell_z^+$ (\ref{eq:lzeff}). The underlying assumption of the linear law, $\Delta U^+ = \ell_x^+ - \ell_z^+$, is that the only effect of the quasi-streamwise vortices is to induce a Couette-like, transverse shear at the reference plane $y=0$, but no wall-normal velocity, as portrayed in figure~\ref{fig:v_vortex}(\textit{a}). This is valid as long as $\ell_z^+ \lesssim 1$, since $w$ is linear just above the wall, whereas $v$ is quadratic, and hence vanishes more rapidly with $y$. In this regime, the effect of the surface on the flow would be captured by the conditions $u=\ell_x \partial u/\partial y$, $w=\ell_z\partial w/\partial y$ and $v = 0$ at $y=0$. This is the regime contemplated by the pioneering work of \citet{Luchini1991}, which is consistent with the homogenisation approaches of~\citet{Lauga2003}, \citet{Kamrin2010}, \citet{Luchini2013} and \citet{Lacis2017}. However, as $\ell_z^+$ increases and the vortices further approach the reference plane, the assumption of impermeability is no longer valid, since, for the vortices to continue to approach the reference plane unimpeded, they would require a non-negligible wall-normal velocity at $y=0$. This concept is depicted in figure~\ref{fig:v_vortex}(\textit{b}).  \citet{Gomez-de-Segura2018a} argue, therefore, that the displacement, on average, of the vortices towards the reference plane would necessarily saturate eventually, unless the shift of the origin perceived by $w$ {was} also accompanied by a corresponding shift of the origin perceived by $v$.  They conducted preliminary {simulations to} find a suitable method {to impose} a virtual origin {on} $v$ {and} to test this hypothesis. {Amongst the several methods studied, \citet{Gomez-de-Segura2020} concluded that a Robin boundary condition at $y=0$, $v=\ell_y \partial v/\partial y$, was a simple yet suitable one.} The inclusion of a wall-normal `slip length' $\ell_y$, sometimes referred to as a `transpiration length',  can be introduced in a homogenisation framework using second-or-higher order expansion~\citep{Bottaro2019,Lacis2020,Bottaro2020}. Irrespective of the texture, and focussing solely on the overlying flow, if the origin perceived by the spanwise and wall-normal velocity fluctuations is the same, \citet{Gomez-de-Segura2018a} observed that the saturation in the effect of $\ell_z^+$ no longer occurred. This suggests that, in general, to fully describe the effects of a small-textured surface on the flow, it may be necessary to consider virtual origins for all three velocity components, because the virtual origin for $v$ can also play an important role in setting the apparent origin for the quasi-streamwise vortices. This implies that, when the virtual origins perceived by $v$ and $w$ differ, the quasi-streamwise vortices, and hence the overlying turbulence, might perceive a virtual origin at some intermediate plane between the two \citep{Gomez-de-Segura2018a,Garcia-Mayoral2019}. This is extensively investigated in \S\,\ref{sec:results}.

\begin{figure}
\centering
\includegraphics[width=0.8\textwidth]{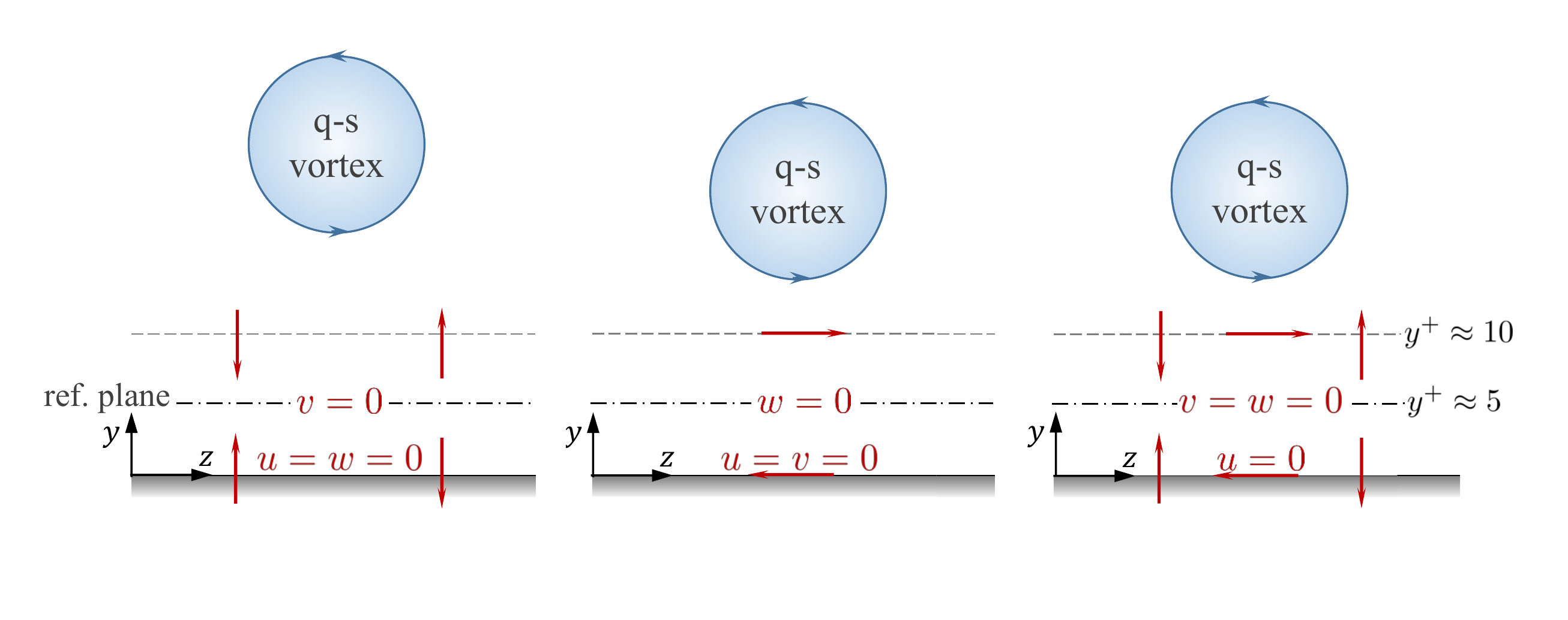}
\mylab{-10.5cm}{3.2cm}{($a$)}%
\mylab{-6.7cm}{3.2cm}{($b$)}%
\mylab{-3.5cm}{3.2cm}{($c$)}%
\caption{{Schematics of the different control strategies studied by \citet{Choi1994}. Opposition control applied on (\textit{a}) $v$, (\textit{b}) $w$, (\textit{c}) both $v$ and $w$. Shaded grey regions indicate the notional virtual origin perceived by the vortex, i.e.\ the virtual origin for turbulence.}}
\label{fig:VOs_choi}
\end{figure}

{As well as textured surfaces that passively impose virtual origins on the three velocity components, \citet{Gomez-de-Segura2018a} discussed the possibility that the effect of active opposition control~\citep{Choi1994} could also be interpreted in terms of virtual origins. This idea stems from the observation that opposition control, when applied to the wall-normal velocity alone, would establish a `virtual wall' approximately halfway between the detection plane, $y^+=y_d^+$, and the physical wall, $y^+=0$ \citep{Hammond1998}.  \citet{Gomez-de-Segura2018a} extended this concept to the general case where all three velocity components could be opposed, which would, in principle, result in each velocity component perceiving a different virtual origin at a plane above the physical wall. \citet{Choi1994} explored several active control strategies, including opposition control of the wall-normal velocity alone ($v$ control), of the spanwise velocity alone ($w$ control), and of both $w$ and $v$ ($w$-$v$ control). Schematics of these three strategies are shown in figure~\ref{fig:VOs_choi}. In each case, the velocity components imposed at the wall, $y^+ = 0$, were opposite to those measured at $y^+ \approx 10$. \citet{Choi1994} reported that the control caused an upward shift of the log law and an outward shift of the turbulence intensities, compared to the uncontrolled flow. These findings are consistent with the reduction in skin friction being a result of an outward shift of the origin perceived by turbulence with respect to the mean flow, which is the same mechanism by which many textured surfaces are understood to reduce drag. In \S\,\ref{sec:opposition}, we explore this idea further by analysing the effect of opposition control on the turbulence statistics in terms of the apparent virtual origins perceived by each velocity component.}

\section{Numerical method}\label{sec:numerics}

\subsection{{Set-up of virtual-origin simulations}}\label{subsec:vo_setup}
{Here we outline the numerical method and summarise the series of simulations with Robin boundary conditions that we carry out.} We conduct direct numerical simulations of turbulent channel flows in a domain doubly periodic in the wall-parallel directions, using a code adapted from \citet{Garcia-Mayoral2011b} and \citet{Fairhall2018}. We solve the non-dimensional, unsteady, incompressible Navier--Stokes equations
\begin{align}
\frac{\partial \boldsymbol{u}}{\partial t} + \boldsymbol{u} \cdot \nabla \boldsymbol{u} &= - \nabla p + \frac{1}{\Rey}\nabla^2 \boldsymbol{u},\\
\nabla \cdot \boldsymbol{u} &= 0,
\end{align}
where $\boldsymbol{u}=(u,v,w)$ is the velocity vector with components in the streamwise, wall-normal and spanwise directions, $x$, $y$ and $z$, respectively, $p$ is the kinematic pressure and $\Rey$ is the channel bulk Reynolds number. {In the streamwise and spanwise directions, the primitive variables are solved in Fourier space,} applying the $2/3$ dealiasing rule when computing the nonlinear advective terms. The wall-normal direction is discretised using a second-order centred finite difference scheme on a staggered grid. Time integration is carried out using a fractional step method \citep{Kim1985}, along with a three-step Runge-Kutta scheme.  The same coefficients as \citet{Le1991} are used, for which semi-implicit and explicit schemes are used to approximate the viscous and advective terms, respectively.

{Simulations are primarily conducted} at friction Reynolds number $\Rey_\tau~=~\delta u_\tau / \nu~\approx~180$. Even though this is a comparatively low Reynolds number, previous studies have shown that it is sufficient to capture the key physics in flows where the effect of surface manipulations is confined to the near-wall region~\citep{Martell2009,Busse2012,Garcia-Mayoral2012}. {Two simulations are also conducted at $\Rey_\tau \simeq 550$ for comparison,} and we will show in \S\,\ref{subsec:reEffects} how our results are scalable to higher Reynolds numbers. {In all cases, the channel half-height is $\delta=1$. For {most simulations,} the wall-parallel domain size is $L_x=2\pi\delta$ and $L_z=\pi\delta$.} This {has been shown by \citet{Lozano-Duran2014}} to be sufficiently large to capture the key turbulence processes and length scales of the near-wall and log-law regions of the flow, and reproduce well the one-point statistics of domains of larger size. { We show that this is the case also here by running one of the simulations at $\Rey_\tau \simeq 550$ with domain size $L_x=8\pi\delta$  and $L_z=3\pi\delta$.} In the wall-parallel directions, the resolution in collocation points is $\Delta x^+ \approx 6$ and $\Delta z^+ \approx 3$ for simulations at $\Rey_\tau \simeq 180$, and $\Delta x^+ \approx 9$ and $\Delta z^+ \approx 4$ for the simulation at $\Rey_\tau \simeq 550$. In the wall-normal direction, the grid is stretched such that $\Delta y^+_{min} \approx 0.3$ at the wall and $\Delta y^+_{max} \approx 3$ at the channel centre. The flow is driven by a constant streamwise pressure gradient, in order to keep $\Rey_\tau$ fixed. The variable time step is controlled by
\begin{equation}
\Delta t = 
\min\left\lbrace 
0.7\left[ 
\frac{\Delta x}{\pi|u|},\frac{\Delta z}{\pi|w|},\frac{\Delta y}{\pi|v|} 
\right],
 2.5\left[ 
\frac{\Delta x^2}{\pi^2}\nu,\frac{\Delta z^2}{\pi^2}\nu,\frac{\Delta y^2}{4}\nu 
\right]
\right\rbrace,
\end{equation}
which corresponds to maintaining a convective CFL number of 0.7 and a viscous one of 2.5. In all cases, the flow was allowed to evolve until any initial transients had decayed, and then statistics were collected over a window of at least 20 largest-eddy turnover times, $\delta/u_\tau$.

Virtual origins for the three velocity components are introduced by imposing Robin, slip-length boundary conditions at the channel walls, following {\citet{Gomez-de-Segura2020}}. {At the bottom wall of the channel, these} take the form
\begin{equation}
u\rvert_{y=0} = \left.\ell_x \frac{\partial u}{\partial y}\right\rvert_{y=0}, \quad v\rvert_{y=0}=  \left. \ell_y \frac{\partial v}{\partial y}\right\rvert_{y=0} \quad \text{and} \quad w\rvert_{y=0}=   \left.\ell_z \frac{\partial w}{\partial y}\right\rvert_{y=0},\label{eq:slip_lengths}
\end{equation} 
so that $u$, $v$ and $w$ at the domain boundary are related to their respective wall-normal gradients by the three slip lengths,  $\ell_x$, $\ell_y$ and $\ell_z$. {Equivalent, symmetric boundary conditions are also applied to the top wall of the channel.} The coupling between velocity components, their wall-normal gradients and the pressure is fully implicit and embedded in the LU factorisation intrinsic in the fractional-step method \citep{Perot1993}. A detailed description of the implementation of this type of boundary conditions can be found in \cite{Gomez-de-Segura2019}. Note that for $v$, as mentioned in \S\,\ref{sec:introduction}, $\ell_y$ does not convey a slip effect, but, by extension, we will also refer to $\ell_y$ as the `slip length' in the wall-normal direction. To prevent any net surface mass flux, the slip length for the $xz$-averaged wall-normal velocity is set to zero, and hence $\ell_y$ is only applied to its fluctuating component. Note also that a free-slip condition, e.g.\ $\partial u/\partial y = 0$, is equivalent, in principle, to imposing an infinitely large slip length, $\ell_x=\infty$.

\begin{figure}
\centering
\vspace{0.1cm}
\includegraphics[scale=0.9]{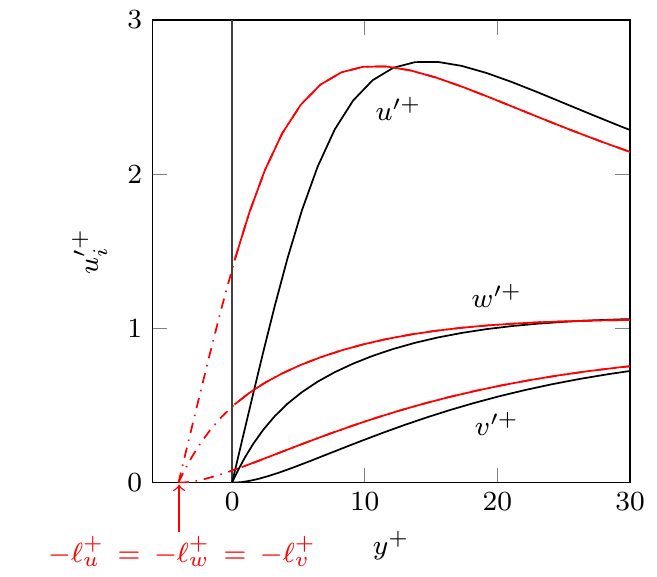}
\hspace{0.5cm}
\includegraphics[scale=0.9]{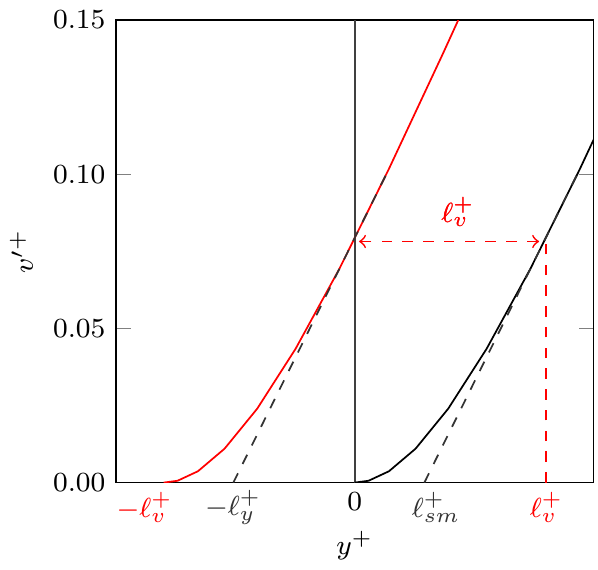}
\mylab{-10.75cm}{4.7cm}{($a$)}%
\mylab{-4.4cm}{4.7cm}{($b$)}%
\caption{Schematics showing ($a$) the definition of virtual origins $\ell_u^+$, $\ell_w^+$ and $\ell_v^+$ as the shift of the r.m.s. velocity fluctuations with respect to a smooth channel; ($b$) the distinction between $\ell_v^+$ and $\ell_y^+$.}
\label{fig:VO_schematic}
\end{figure}

\begin{table}
\begin{center}
\begin{tabular}{l|ccc|cccc|ccc|ccc}
Case & $Re_{\tau}$ &{$L_x/\delta$}& {$L_z/\delta$}& $\ell_x^+$ & $\ell_z^+$ & $\ell_y^+$ & $\ell_{x,m}^+$ & $\ell_u^+$ & $\ell_w^+$ & $\ell_v^+$ & $\ell_U^+$ & $\ell_T^+$ & $\ell_{T,pred}^+$ \\[6pt]
V1   & 180     &$2\pi$&$\pi$     & 0.0        & 0.0        & 1.2        & -              & 0.0        & 0.0        & 1.9        & 0.0        & 0.0        & 0.0               \\
V2   & 180     &$2\pi$&$\pi$    & 0.0        & 0.0        & 2.5        & -              & 0.0        & 0.0        & 3.9        & 0.0        & 0.0        & 0.0               \\
UV1  & 180    &$2\pi$& $\pi$      & 2.0        & 0.0        & 1.2        & -              & 2.0        & 0.0        & 1.9        & 2.0        & 0.0        & 0.0               \\
UV2  & 180    &$2\pi$&  $\pi$    & 4.0        & 0.0        & 2.5        & -              & 3.6        & 0.0        & 3.9        & 4.0        & 0.0        & 0.0             \\[6pt]
UM1  & 180   &$2\pi$&   $\pi$     & 5.0        & 0.0        & 0.0        & 0.0            & 4.3        & 0.0        & 0.0        & 0.0        & 0.0        & 0.0               \\
UM2  & 180   &$2\pi$&$\pi$        & 10.0       & 0.0        & 0.0        & 0.0            & 6.6        & 0.0        & 0.0        & 0.0        & 0.0        & 0.0               \\
UM3  & 180   &$2\pi$& $\pi$       & 20.0       & 0.0        & 0.0        & 0.0            & 8.8        & 0.0        & 0.0        & 0.0        & 0.0        & 0.0               \\
UM4  & 180   &$2\pi$&$\pi$        & 50.0       & 0.0        & 0.0        & 0.0            & 11.2       & 0.0        & 0.0        & 0.0        & 0.0        & 0.0               \\
UM5  & 180   &$2\pi$& $\pi$       & 100.0      & 0.0        & 0.0        & 0.0            & 12.6       & 0.0        & 0.0        & 0.0        & 0.0        & 0.0               \\
UM6  & 180   &$2\pi$& $\pi$       & $\infty$   & 0.0        & 0.0        & 0.0            & $\infty$   & 0.0        & 0.0        & 0.0        & 0.0        & 0.0               \\[6pt]
UWV1 & 180  &$2\pi$& $\pi$        & 2.0        & 2.0        & 1.2        & -              & 2.0        & 1.7        & 1.9        & 2.0        & 1.7        & 1.7               \\
UWV2 & 180  &$2\pi$& $\pi$        & 3.0        & 3.0        & 1.9        & -              & 2.9        & 2.4        & 3.0        & 3.0        & 2.4        & 2.4               \\
UWV3 & 180  &$2\pi$& $\pi$        & 4.0        & 4.0        & 1.2        & -              & 3.6        & 2.9        & 1.9        & 4.0        & 2.7        & 2.7               \\
UWV3H & 550&$2\pi$& $\pi$          & 4.0        & 4.0        & 1.2        & -              & 3.6        & 2.9        & 1.9        & 3.9        & 2.6        & 2.7               \\
{UWV3HD} & 550  &$8\pi$&$3\pi$         & 4.0        & 4.0        & 1.2        & -              & 3.6        & 2.9        & 1.9        & 3.9        & 2.6        & 2.7               \\
UWV4 & 180 &$2\pi$&$\pi$          & 4.0        & 4.0        & 2.5        & -              & 3.6        & 2.9        & 3.9        & 3.8        & 2.9        & 2.9               \\
UWV5 & 180  &$2\pi$&$\pi$         & 4.0        & 6.0        & 1.5        & -              & 3.6        & 3.9        & 2.3        & 3.9        & 3.5        & 3.5               \\
UWV6 & 180   &$2\pi$&$\pi$        & 4.0        & 6.0        & 2.0        & -              & 3.6        & 3.9        & 3.2        & 3.8        & 3.8        & 3.8               \\
UWV6M & 180 &$2\pi$&$\pi$          & 4.0        & 6.0        & 2.0        & 10.0           & 3.6        & 3.9        & 3.2        & 9.4        & 3.8        & 3.8               \\[6pt]
UWVL1  & 180  &$2\pi$&$\pi$         & 6.0        & 2.0        & 4.5        & -              & 4.9        & 1.7        & 6.3        & 5.7        & 2.4        & 1.7               \\
UWVL2  & 180  &$2\pi$&$\pi$         & 5.0        & 5.0        & 5.0        & -              & 4.3        & 3.4        & 6.8        & 4.4        & 4.7        & 3.4               \\
UWVL3  & 180  &$2\pi$&$\pi$         & 8.2        & 11.0       & 4.3        & -              & 5.9        & 5.9        & 6.0        & 6.7        & 5.7        & 5.9               \\
UWVL4  & 180 &$2\pi$&$\pi$          & 10.0       & 10.0       & 10.0       & -              & 6.6        & 5.5        & 11.2       & 6.0        & 7.7        & 5.5   \\[6pt]
WV1  & 180  &$2\pi$&$\pi$         & 0.0        & 2.0        & 1.2        & -              & 0.0        & 1.7        & 1.9        & 0.0        & 2.1        & 1.7               \\
WV2  & 180  &$2\pi$&$\pi$         & 0.0        & 4.0        & 2.5        & -              & 0.0        & 2.9        & 3.9        & 0.0        & 4.4        & 2.9               \\
WV3  & 180  &$2\pi$&$\pi$         & 0.0        & 6.0        & 2.2        & -              & 0.0        & 3.9        & 3.4        & 0.0        & 4.9        & 3.9                         
\end{tabular}
\caption{Summary of simulations, including the slip lengths used for the boundary conditions, $\ell_x^+$, $\ell_z^+$ and $\ell_y^+$, and their corresponding virtual origins, $\ell_u^+$, $\ell_w^+$ and $\ell_v^+$, calculated a priori from the smooth-wall profiles. The slip length for the mean flow, $\ell_{x,m}^+$, is given only when it is different to the slip length for the streamwise velocity fluctuations. Note that, here, $\Rey_{\tau}$ is the friction Reynolds number calculated with respect to the plane $y=0$. The virtual origin for the mean flow, $\ell_U^+$, is given as the mean streamwise slip velocity, $U^+_s$, measured at $y=0$. The virtual origin for turbulence, $\ell_T^+$, is found a posteriori and compared to that predicted by equation (\ref{eq:lTp}), $\ell^+_{T,pred}$. In the case names, `U', `V' and `W' denote a non-zero slip-length boundary condition on $u$, $v$ and $w$, respectively, `M' signifies that the slip applied to the streamwise velocity fluctuations is not the same as that applied to {(M)}ean velocity,  {`H' is for the {(H)}igher Reynolds number cases at $\Rey_{\tau}=550$, `D' is for the simulation with the larger {(D)}omain in the streamwise and spanwise directions,} and `L' is for cases with {(L)}arge slip lengths. Note that the slip lengths,  $\ell_x^+$, $\ell_z^+$ and $\ell_y^+$, and virtual origins, $\ell_u^+$, $\ell_w^+$ and $\ell_v^+$, are scaled with the friction velocity measured at the domain boundary, $y=0$, whereas $\ell_U^+$ and $\ell_T^+$ are scaled with the friction velocity measured at the origin for turbulence $y=-\ell_T^+$. The origin for turbulence predicted from (\ref{eq:lTp}), $\ell_{T,pred}^+$, is scaled with the friction velocity at that origin,  i.e. at $y=-\ell_{T,pred}^+$.}
\label{tab:slipLengthsVOs}
\end{center}
\end{table}

While the concepts of slip lengths and virtual origins have been used interchangeably in the literature, here we make a subtle but important difference. We will denote by $\ell_x^+$, $\ell_y^+$ and $\ell_z^+$ the slip lengths in the streamwise, wall-normal and spanwise directions, respectively, which are defined exclusively as the Robin coefficients for the simulation boundary conditions~(\ref{eq:slip_lengths}). Physically, they simply correspond to the wall-normal locations where the velocity components become zero when linearly extrapolated from the reference plane, $y=0$. {In order to associate the imposed slip lengths with smooth-wall data a priori, we define the virtual origins of $u$, $v$ and $w$ as the notional distance below the reference plane where each velocity component would perceive a virtual, smooth wall. To do this, we assume the shape of each r.m.s.-fluctuation profile {would remain} the same as over a smooth wall, independently of the others. The virtual origins would then be located at $y^+=-\ell_u^+$, $y^+=-\ell_v^+$ and $y^+=-\ell_w^+$, respectively. We note that {this is not physics-based, but it simply}  allows us to establish an a priori correspondence between the offset in each velocity component and the slip length for that velocity {while accounting for the non-linear behaviour of the fluctuating velocities, especially for $v$, near the wall.}} The definition of these virtual origins is illustrated in figure~\ref{fig:VO_schematic}($a$). The slip lengths for the Robin boundary conditions~(\ref{eq:slip_lengths}) are therefore set with the objective of yielding a prescribed set of virtual origins $\ell_u$, $\ell_v$ and $\ell_w$. Table~\ref{tab:slipLengthsVOs} summarises the parameters of the simulations that we conduct in this study. For each case, the slip lengths $\ell_x^+$, $\ell_z^+$ and $\ell_y^+$ are given, along with the corresponding virtual origins $\ell_u^+$, $\ell_w^+$ and $\ell_v^+$. Since the virtual origins are computed from the slip lengths a priori, assuming the shape of smooth-wall velocity profiles remain unchanged, there is a one-to-one a priori relationship between slip lengths and virtual origins. {The simulations are split into various ‘families’, each designed to systematically test a particular aspect of this virtual-origin framework. For example, some cases impose a virtual origin on $v$ alone (denoted by ‘V’), while other cases impose a virtual origin on both $u$ and $v$ (denoted by ‘UV’), and so on. The exact purpose of each family is explained in  \S\,\ref{sec:results}. Note that for} some of the simulations, the slip length applied to the mean flow, $\ell_{x,m}$, is different to the slip length applied to the streamwise fluctuations, $\ell_{x}$. Since we solve the flow in Fourier space in the wall-parallel directions, this can be implemented easily by imposing different slip-length boundary conditions on the different modes $\hat{u}(k_x,k_z,y)$ as required, where $k_x$ and $k_z$ are the streamwise and spanwise wavenumbers, respectively. 

For a virtual origin of a few wall units, we expect the slip lengths $\ell_x^+$ and $\ell_z^+$ to be approximately equal to $\ell_u^+$ and $\ell_w^+$, because the wall-parallel velocities $u^+$ and $w^+$ are essentially linear in the immediate vicinity of the wall. The case of the wall-normal velocity, however, is less straightforward. Since $v^{\prime+}$ is essentially quadratic very near the wall, the height of the virtual origin perceived by $v^+$, $y^+=-\ell_v^+$, can differ significantly from the slip length $\ell_y^+$, even for small values, as illustrated in figure~\ref{fig:VO_schematic}($b$). We choose $\ell_y^+$ as the ratio between $v^{\prime+}$ and $\mathrm{d} v^{\prime+} /\mathrm{d} y^+$ at a height $y^+=\ell_v^+$ above a smooth wall. From figure~\ref{fig:VO_schematic}($b$), $\ell_y^+$ and $\ell_v^+$ are related by $\ell_y^+=\ell_v^+-\ell_{sm}^+$, where $\ell_{sm}^+$ is obtained by linearly extrapolating the slope of the smooth-wall profile at $y^+=\ell_v^+$. Note that the value of $\ell_{sm}^+$ is a function of $\ell_v^+$, as it depends on the local slope of the profile at the height from which the extrapolation is calculated. A curvature effect can also be significant for $\ell_w^+$, since the profile of $w^{\prime +}$ becomes noticeably curved for $y^+\gtrsim 2$, but this effect is small for $u^{\prime+}$.  Since the mean velocity profile is approximately linear up to $y^+ \approx 5$, the distance below the plane $y^+=0$ of the virtual origin experienced by the mean flow, $\ell_U^+$, is essentially equal to the slip velocity of the mean flow in wall units, $U_s^+$, and also to its slip length, $\ell_{x,m}^+$. It should, however, be mentioned that in general, if $\ell_x^+$ is large enough, the virtual origin perceived by the mean flow, $y^+ = -\ell_U^+$, is not necessarily coincident with the virtual origin for the streamwise fluctuations, $y^+=-\ell_u^+$, since their profiles curve differently as they approach the wall, even if $\ell_x^+ = \ell_{x,m}^+$.

\subsection{{Set-up of opposition-control simulations}}\label{subsec:oc_setup}

{As well as the virtual-origin simulations {described above}, we also investigate the effect of opposition control~\citep{Choi1994} from the viewpoint of virtual origins. We carry out three simulations at $\Rey_\tau \approx 180$, applying opposition control to $v$ alone, $w$ alone, and both $v$ and $w$. The same DNS code as {for the virtual-origin simulations is used, with the only difference being the imposed boundary conditions.} The control is implemented in the code explicitly, with the measured velocity at the plane $y^+=y_d^+$ at time step $n$ is opposed at the wall at time step $n + 1$. The detection plane is set at $y_d^+ = 7.8$, with the aim of generating notional virtual origins for the controlled velocities at $y^+\approx 4$, similar to our virtual-origin simulation UWV6. A summary of the opposition-control simulations is given in table~\ref{tab:opposition}, including  several parameters relevant to their interpretation in terms of virtual origins, which will be discussed in detail in \S\,\ref{sec:opposition}.}

\begin{table}
\begin{center}
\begin{tabular}{lccccccc}
Case & $y_d^+/2$ & $\ell_U^+$ & $\ell_w^+$ & $\ell_v^+$ & $\ell_{T,pred}^+$&$\ell_U^+ - \ell_{T,pred}^+$& $\Delta U^+$\\[6pt] 
$v$ control          & 3.9 & {0.0} & {$\hphantom{-}0.0$} & {$-3.9$} & {$-1.7$} & 1.7 & 1.9\\          
$w$ control         & 3.9 & {0.0}& {$-3.9$} &{$\hphantom{-}0.0$} & {$-3.9$}& 3.9 & 3.0\\
$w$-$v$ control & 3.9 & {0.0}& {$-3.9$} & {$-3.9$} & {$-3.9$} & 3.9 & 3.7
\end{tabular}
\caption{{Summary of opposition control simulations. For each case, the notional virtual origins are given with respect to the reference plane $y^+=0$, assuming that the control establishes a virtual origin for the opposed velocity components at $y^+=y_d^+/2$, where $y_d^+$ is the detection plane height. The predicted virtual origin for turbulence, $\ell_{T,pred}^+$, is given, which is calculated from (\ref{eq:lTp}). The difference $\ell_U^+ - \ell_{T,pred}$ represents the predicted shift in the mean velocity profile, and $\Delta U^+$ is the measured shift in the mean velocity profile from figure~\ref{fig:opposition}.}}
\label{tab:opposition}
\end{center}
\end{table}

\section{{Analysis of virtual-origin simulations}}\label{sec:results}

In this section, we discuss the results of the DNSs with Robin slip-length boundary conditions (\ref{eq:slip_lengths}) summarised in table~\ref{tab:slipLengthsVOs}. The aim is to determine the effect on the flow of imposing different virtual origins for each velocity component. In particular, we are concerned with how $\Delta U^+$ and the near-wall turbulence dynamics are affected by the virtual origins. We also wish to better understand the physical mechanism at play, such that we can potentially predict the effect of the virtual origins on the flow a priori.

\subsection{The origin for turbulence}\label{subsec:originForTurb}

{In \S\,\ref{sec:theory}, we introduced} the idea that the quasi-streamwise vortices, and hence the turbulence, might perceive an intermediate origin between the virtual origins perceived by $v$ and $w$. We now discuss this concept in more detail. {Let us postulate that the only effect of the virtual origins, particularly those perceived by $v$ and $w$, on the near-wall turbulence is to set its origin at some intermediate plane, while the flow remains otherwise the same as over a smooth wall. In this paper, we define $\ell_T^+$ as the distance between the virtual origin perceived by turbulence and the reference plane $y^+=0$. When $\ell_T^+ >0$, the virtual origin perceived by turbulence is below the reference plane, {and} therefore we refer to the plane $y^+ = -\ell_T^+$ as the virtual origin for turbulence. Likewise, we denote by $y^+=-\ell_U^+$ the virtual origin perceived by the mean flow.} It follows from the streamwise momentum equation that the {shape of the} mean velocity profile {in a channel} is determined by the turbulence through the Reynolds stress~{\citep{Pope2000,Gomez-de-Segura2020}}. If $\ell_U^+ > \ell_T^+$, the virtual origin perceived by the mean flow is deeper than that perceived by the turbulence. {In this case, the mean velocity profile would be free to grow with essentially unit gradient in wall units from $y^+=-\ell_U^+$ to $y^+=-\ell_T^+$, due to the absence of Reynolds shear stress in the region $-\ell_U^+\leq y^+\leq -\ell_T^+$.}  Above $y^+=-\ell_T^+$, the Reynolds {shear stress would be the same as over a smooth wall, and so would the shape of the mean velocity profile, but shifted by} the additional velocity $U^+(y^+=-\ell_T^+) = \ell_U^+ - \ell_T^+$. Note that the above ideas apply to the virtual profile that would extend below $y^+=0$, as mentioned {in} \S\,\ref{sec:theory}. {The} outward shift of the mean velocity profile would then necessarily be given by
\begin{equation}
\Delta U^+ = \ell_U^+ - \ell_T^+, \label{eq:DUlT}
\end{equation}
which would propagate to all heights above the plane $y^+=-\ell_T^+$~\citep{Gomez-de-Segura2018a,Garcia-Mayoral2019}. 

{The physical idea described by (\ref{eq:DUlT}) was, in fact, essentially proposed by \citet{Luchini1996}, who postulated that the log law would be modified by the presence of texture only through a shift $\Delta U^+$ ``if the structure of the turbulent eddies were unaltered in the reference frame that has the transverse equivalent wall as origin, whereas the mean flow profile obviously starts at the longitudinal equivalent wall.''} In other words, $\Delta U^+$ should be the height difference between the origin for the mean flow, at $y^+=-\ell_U^+$, and the origin for turbulence, at $y^+=-\ell_T^+$. In this framework, from the point of view of turbulence the `wall' is located at $y^+=-\ell_T^+$, which, therefore, should also be the height of reference when comparing with smooth-wall data. {Note that} (\ref{eq:DUlT}) is based on the assumption that the effect of the texture on the mean flow and the turbulence is only to change the virtual origins that they perceive, and that the dynamics of the near-wall cycle is unaffected. This requires that the flow perceives the surface in a homogenised fashion, and the direct, granular effect of the texture is negligible \citep{Garcia-Mayoral2019}. {In the context of} superhydrophobic  surfaces, {for instance,} \citet{Fairhall2019} show that this is the case so long as the characteristic length scale of the texture {in wall units} satisfies $L^+ \lesssim 25$.  Using the results from our DNSs, we will now examine the validity of (\ref{eq:DUlT}), starting first with the dependence of $\ell_T^+$ on the virtual origins imposed on the three velocity components, $\ell_u^+$, $\ell_v^+$ and $\ell_w^+$. 

{In \S\,\ref{sec:theory} we have discussed} the idea that the quasi-streamwise vortices of the {near-wall cycle} induce, as a first-order effect, a spanwise flow very near the wall and, as a second-order effect, a wall-normal one. This would explain the saturation in the effect of the spanwise slip length, $\ell_z^+$, in the absence of permeability, i.e. when $\ell_y^+=0$. Furthermore, this is consistent with the idea that when the virtual origin perceived by the wall-normal velocity is roughly at the same depth as that perceived by the spanwise velocity, $\ell_v^+ \approx \ell_w^+$, no saturation is observed~\citep{Gomez-de-Segura2018a}. {This implies} that when imposing a virtual origin on the wall-normal velocity alone, without any spanwise slip, the virtual origin perceived by the vortices should remain at the domain boundary,  $y=0$, regardless of { how large $\ell_v^+$ was.} Since, in this case, $w=0$ at the reference plane, and the vortices induce predominantly a spanwise flow in the vicinity of the wall, transpiration alone would not allow the vortices to move any closer to the reference plane. {This is the} contrasting, but complementary concept to the saturation in the effect of $\ell_z^+$ in the absence of transpiration depicted in figure~\ref{fig:v_vortex}.

\begin{figure}
\centering
\includegraphics{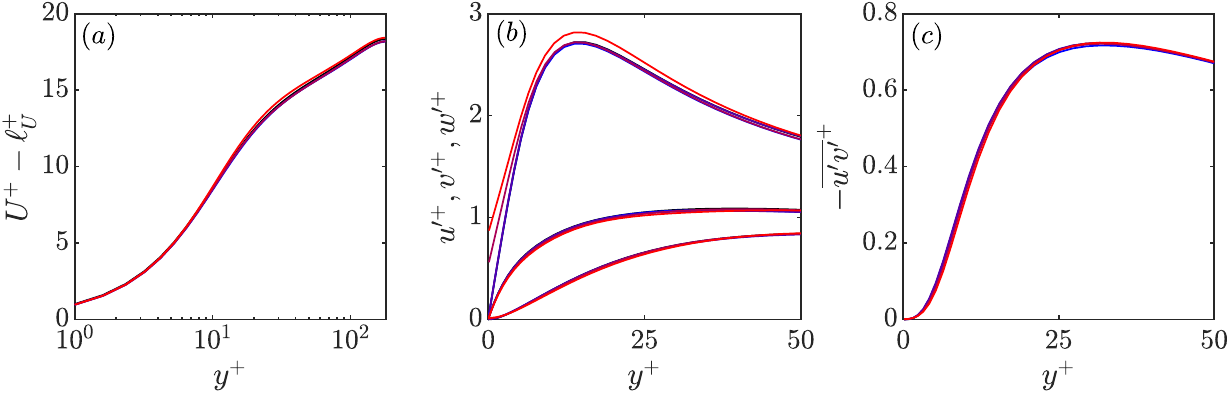}
\caption{Mean velocity profiles, r.m.s. velocity fluctuations and Reynolds shear stress profiles for slip-length simulations with no spanwise slip. Black lines, smooth-wall reference data; blue to red lines, cases V1, V2, UV1 and UV2. Note that in (\textit{a}), the mean streamwise slip length, $\ell_U^+$, where appropriate, has been subtracted from the mean velocity profile.}
\label{fig:no_w}
\end{figure}

{We assess the idea presented in the previous paragraph in simulations V1, V2, UV1 and UV2, all of which have $\ell_w^+=0$.} The mean velocity profiles, r.m.s.\ velocity fluctuations and Reynolds stress profiles for {these simulations} are shown in figure~\ref{fig:no_w}. The figure supports the idea that the virtual origin experienced by the spanwise flow is, indeed, the most limiting in terms of setting the virtual origin for turbulence, and $\ell_T^+=0$ for all cases. For the two cases with a non-zero virtual origin for $v$ only, cases V1 and V2, there is no change in the statistics whatsoever with respect to the smooth-wall data, even for virtual origins as large as $\ell_v^+\approx 4$. When a non-zero virtual origin is also applied to the streamwise flow, such that $\ell_u^+,\ell_v^+ > 0$ but $\ell_w^+=0$, there is still no change in the wall-normal and spanwise r.m.s. velocity fluctuations, $v^{\prime+}$ and $w^{\prime+}$, or the Reynolds stress profile. We also observe that the mean velocity profile is essentially identical to the smooth-wall case, save for the shift $\ell_U^+ = U^+_s$, as shown in figure~\ref{fig:no_w}(\textit{a}). However, the peak value of the streamwise r.m.s. velocity fluctuations, $u^{\prime+}$, increases as $\ell_u^+$ is increased, and the $u^{\prime+}$ curve does not fit well the smooth-wall data near the wall for case UV2, when $\ell_u^+ = 4$. This appears to occur independently of the mean flow and other statistics, and this will be investigated further in \S\,\ref{subsec:streamwiseOrigin}.

\begin{figure}
\centering
\includegraphics{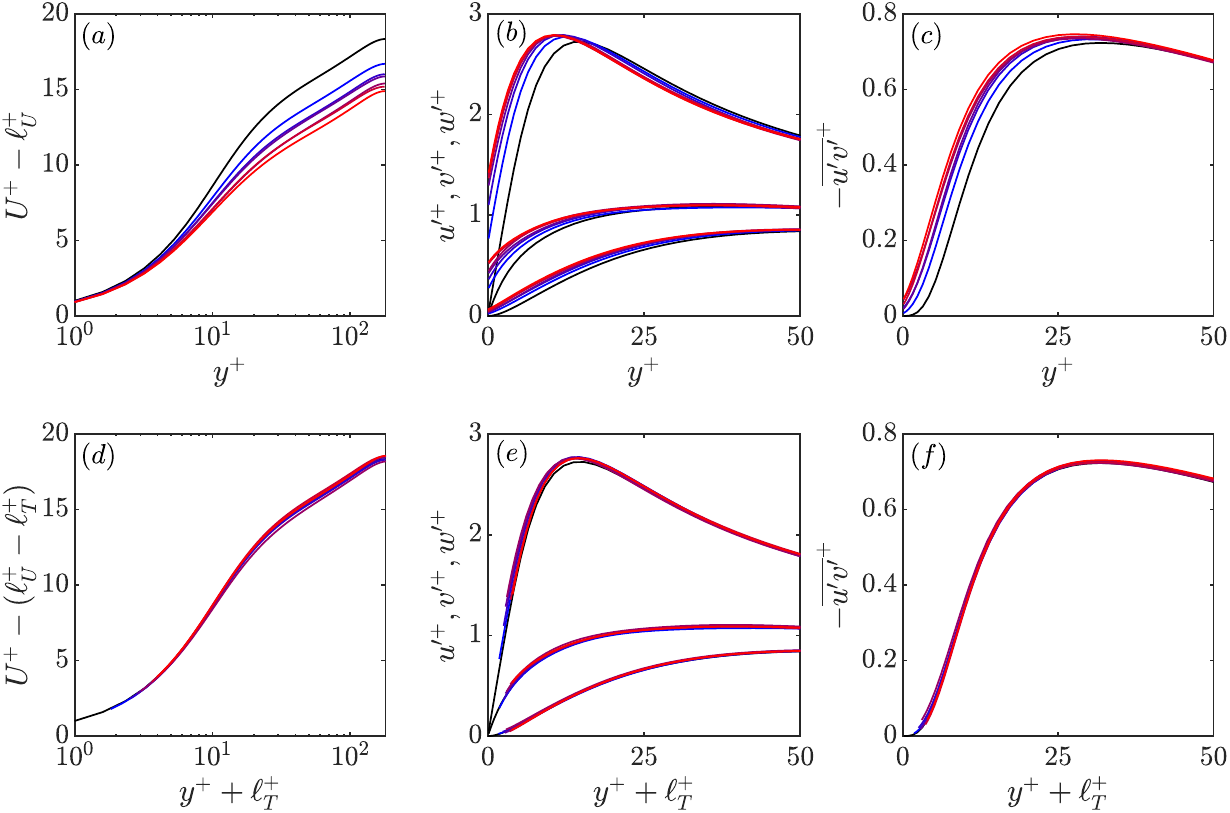}
\caption{Mean velocity profiles, r.m.s. velocity fluctuations and Reynolds shear stress profiles for simulations with non-zero slip-length boundary conditions applied to all three velocity components. (\textit{a}--\textit{c}) scaled with the friction velocity at the reference plane, $y^+=0$; (\textit{d}--\textit{f}) shifted in $y^+$ by $\ell_T^+$ and scaled with the friction velocity at the origin for turbulence, $y^+=-\ell_T^+$. Black lines, smooth-wall reference data; blue to red lines, cases UWV1--UWV6. }
\label{fig:comp_uwv}
\end{figure}

The results presented so far suggest that the quasi-streamwise vortices cannot perceive an origin deeper than the origin perceived by the spanwise velocity. {We now investigate the effect on the flow when both $\ell_w^+$ and $\ell_v^+$ are non-zero, using simulations UWV1--UWV6.} The mean velocity profiles, r.m.s. velocity fluctuations and Reynolds shear stress profiles for {these simulations} are included in figure~\ref{fig:comp_uwv}. These simulations have non-zero slip-length coefficients for all three velocity components, {with} $\ell_x^+\lesssim 4$, $\ell_y^+\lesssim 2$ and $\ell_z^+\lesssim 6$. In figure~\ref{fig:comp_uwv}(\textit{a}), after subtracting $\ell_U^+$ from the mean flow in each case, we see that there is still a noticeable difference between the mean velocity profile of the smooth-wall and the slip-length simulations. This difference is consistent with the origin for turbulence lying below the reference plane, $y^+=0$, which acts to increase the drag. We also observe in figure~\ref{fig:comp_uwv}(\textit{b},\textit{c}) that the velocity fluctuations and Reynolds stress profile are shifted towards $y^+=0$ and that their qualitative shape appears to have changed. 

{The above is the conventional way of representing turbulence statistics in the flow-control literature. For example, the observed reduction in velocity and vorticity fluctuations above riblets has been interpreted by some authors as the quasi-streamwise vortices being modified or damped, as well as the spanwise motion of the near-wall streaks being inhibited~\citep[see e.g.][]{Choi1993,Chu1993,El-Samni2007}. Similarly, in studies on the effects of superhydrophobic surfaces, authors have reported that turbulent structures are weakened, modified or disrupted by the presence of the surface~\citep[see e.g.][]{Min2004,Busse2012,Park2013,Jelly2014}. These interpretations would suggest that the turbulence is no longer as it would be over a smooth wall. However, following the physical arguments leading to (\ref{eq:DUlT}), if the turbulence remains otherwise as it would over a smooth surface, it should be possible to account for the difference with smooth-wall data by a mere origin offset.}

{First we discuss} the choice of the friction velocity $u_\tau$. As mentioned above, if as proposed by \citet{Luchini1996} turbulence perceives a virtual smooth wall at $y = -\ell_T$, it follows that the friction velocity $u_\tau$ that scales the flow would be provided by the shear stress at that height. Since the total stress in a channel is linear with $y$, the friction velocity at $y=-\ell_T$ can be found by simply extrapolating the total stress curve from the domain boundary, $y=0$. This would be given by
\begin{equation}
u_\tau(y=-\ell_T) = u_{\tau,0}\sqrt{\frac{\delta + \ell_T}{\delta}},\label{eq:utau}
\end{equation}
where $u_{\tau,0}$ is the friction velocity measured at $y=0$. Note that the friction velocity measured from the surface drag is not necessarily the same as the friction velocity that sets the scaling for the turbulence. Nevertheless, from (\ref{eq:utau}), the ratio $u_\tau/u_{\tau,0}$ is close to unity as long as $\ell_T/\delta \ll 1$, which will be the case at the typical Reynolds numbers of experiments and engineering applications. As we will see below, even in the cases presented in this study, which are conducted at $\Rey_\tau = 180$, $u_\tau$ measured at $y=-\ell_T$ is never more than about 2\% larger than $u_{\tau,0}$.

Using the friction velocity of equation (\ref{eq:utau}), we can recalculate the viscous length scale and renormalise the measured velocities and Reynolds stress. These profiles can then be shifted in $y^+$ by $\ell_T^+$, where the `+' superscript now indicates scaling in wall units based on the $u_\tau$ computed from (\ref{eq:utau}). {If the turbulence dynamics are indeed unmodified compared to the flow over a smooth wall}, except for this shift in origin, which affects both the wall-normal coordinate and the scaling of the flow, then the r.m.s. velocity fluctuations and Reynolds stress profile should essentially collapse to the smooth-wall data. Since $\Delta U^+ = \ell_U^+ - \ell_T^+$, the only difference between the curve $U^+-\ell_U^+$ and the smooth-wall mean velocity profile should be $\ell_T^+$ at all heights.

{We measure} the virtual origin for turbulence a posteriori in cases UWV1--UWV6 by finding the shift that best fits the Reynolds stress curve to smooth-wall data in the near-wall region, $5 \lesssim y^+ + \ell_T^+ \lesssim 20$, and compute the friction velocity at this origin from (\ref{eq:utau}). The measured value of $\ell_T^+$ is included in table~\ref{tab:slipLengthsVOs} for each case, along with the value for all the other cases considered in this study. The figure shows that when the wall-normal coordinate is measured from the virtual origin for turbulence, {$y^+=-\ell_T^+$}, the wall-normal and spanwise r.m.s. fluctuations and Reynolds stress curves essentially collapse to the smooth-wall data, as shown in figures~\ref{fig:comp_uwv}(\textit{e},\textit{f}). {This suggests that, in these cases, the turbulence remains essentially unchanged compared to the flow over a smooth wall. We will refer to this as the turbulence being essentially `smooth-wall like'. Further, this implies that any apparent modifications {to turbulence} that might be concluded from figures~\ref{fig:comp_uwv}(\textit{b},\textit{c}) are actually an apparent effect {caused} by the way the data is portrayed.} Let us note that the resulting $v^{\prime+}$ and $w^{\prime+}$ profiles appear to perceive an origin at $y^+ = - \ell_T^+$, and not the ones prescribed a priori, $y^+ = - \ell_v^+$ and $y^+ = - \ell_w^+$. This is the expected result if $v^{\prime+}$ and $w^{\prime+}$ arise from smooth-like near-wall dynamics and are {thus intrinsically coupled.} The offsets $\ell_v^+$ and $\ell_w^+$ are merely prescribed, a priori values to quantify the offset in $v$ and $w$ caused by the surface, but turbulence would react to their combined effect, perceiving a single origin if it is to remain smooth-wall-like. There are some small deviations from the smooth-wall data for $u^{\prime+}$, which will be discussed in \S\,\ref{subsec:streamwiseOrigin}. Significantly, figure~\ref{fig:comp_uwv}(\textit{d}) demonstrates that the mean velocity profile is also smooth-wall-like, when plotted against $y^++\ell_T^+$, save for the difference $\Delta U^+ =  \ell_U^+-\ell_T^+$. This strongly supports the validity of (\ref{eq:DUlT}). 

\begin{figure}
\centering
\includegraphics{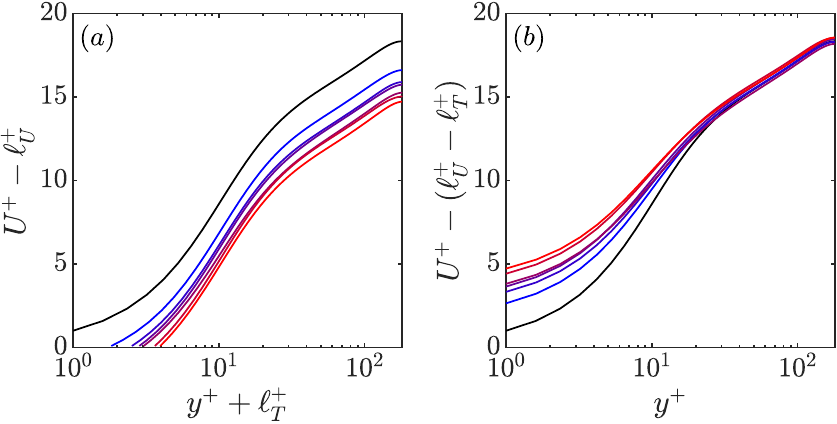}
\caption{Mean velocity profiles for cases UWV1--UWV6, scaled with the friction velocity at the origin for turbulence, $y^+ = -\ell_T^+$: (\textit{a}) $U^+ - \ell_U^+$ with the wall-normal coordinate measured from the origin for turbulence,  $y^+ = -\ell_T^+$; (\textit{b}) $U^+ - (\ell_U^+-\ell_T^+)$ with the wall-normal coordinate measured from the boundary, $y^+=0$. Black lines, smooth-wall reference data; blue to red lines, cases UWV1--UWV6. }
\label{fig:comp_uwv_offsetTest}
\end{figure}

{For comparison}, two other possible ways of portraying the mean velocity profiles for cases UWV1--UWV6 are included in figure~\ref{fig:comp_uwv_offsetTest}. Once the friction velocity is computed at the origin for turbulence, $y^+=-\ell_T^+$, and the wall-normal coordinate is also measured from that height, the mean velocity profiles from the slip-length simulations are essentially parallel to the smooth-wall one for all $y^+$, as shown in figure~\ref{fig:comp_uwv_offsetTest}(\textit{a}). The only difference between the curves of $U^+ - \ell_U^+$ plotted against $y^++\ell_T^+$ from the slip-length simulations and the smooth-wall mean velocity profile is the origin for turbulence, $\ell_T^+$, as mentioned above. Alternatively, again computing the friction velocity at $y^+=-\ell_T^+$, but now leaving $y^+=0$ as the datum for the wall-normal coordinate, the profiles of $U^+ - \Delta U^+$ collapse to the smooth-wall profile only for $y^+\gg1$, as portrayed in figure~\ref{fig:comp_uwv_offsetTest}(\textit{b}). In other words, the profiles collapse to the smooth-wall data only above the near-wall region of the flow \citep{Clauser1956}. The choice of axes in figure~\ref{fig:comp_uwv_offsetTest}(\textit{b}) would indicate that we have measured the correct $\Delta U^+$, but would not suggest that the profiles are smooth-wall-like across the whole $y^+$ range. The only way that they will collapse immediately from $y^+ = 0$ is to measure the wall-normal coordinate from the plane $y^+=-\ell_T^+$, as already shown in figure~\ref{fig:comp_uwv}(\textit{d}). This also emphasises the idea that equation (\ref{eq:DUlT}) will only hold if the origin for turbulence, i.e.\ the plane $y^+=-\ell_T^+$, is used as reference for the turbulence dynamics, setting their scaling for velocity and length, as well as their height origin.

\begin{figure}
\centering
\includegraphics{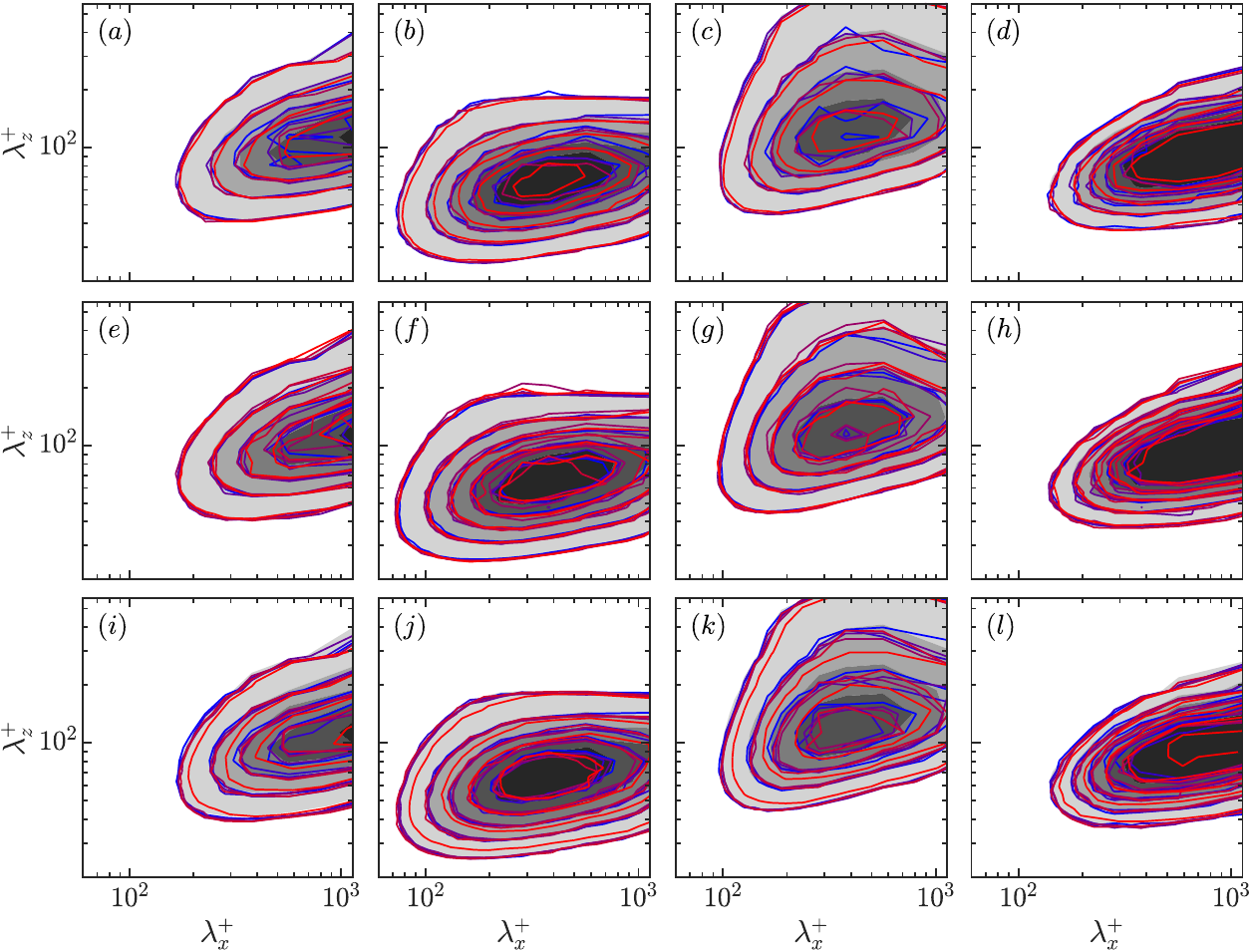}
\caption{Premultiplied two-dimensional spectral densities of $u^2$, $v^2$, $w^2$ and $uv$ at $y^+ + \ell_T^+ = 15$, normalised by $u_\tau$ at the origin for turbulence, $y^+ = -\ell_T^+$, for various slip-length simulations (line contours), compared to smooth-wall data (filled contours) at $y^+ = 15$. (\textit{a}--\textit{d}) cases V1, V2, UV1 and UV2, with line colours as in figure~\ref{fig:no_w}. {(\textit{e}--\textit{h}) cases UWV1--UWV6, with line colours as in figure~\ref{fig:comp_uwv}.  (\textit{i}--\textit{l}) cases UM1--UM6, with line colours as in figure~\ref{fig:uprime_only}.}  First column, $(k_x k_z E_{uu})^+$; second column, $(k_x k_z E_{vv})^+$; third column, $(k_x k_z E_{ww})^+$; fourth column, $-(k_x k_z E_{uv})^+$.  The contour increments for each column are 0.3224, 0.0084, 0.0385 and 0.0241, respectively.}
\label{fig:comp_spectra}
\end{figure}

\begin{figure}
\centering
\includegraphics{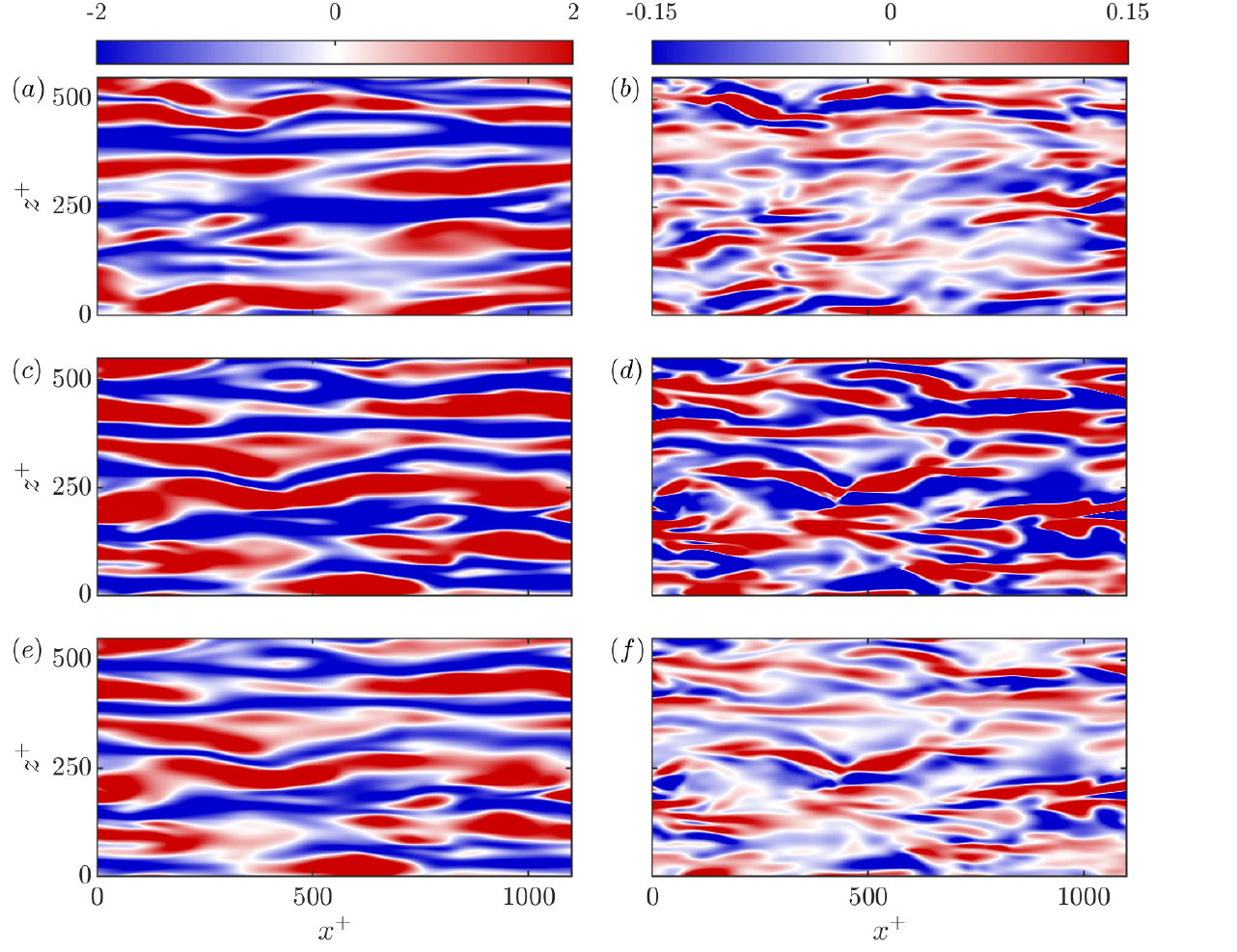}
\caption{Streamwise (\textit{a},\textit{c},\textit{e}) and wall-normal (\textit{b},\textit{d},\textit{f}) instantaneous velocity fluctuation flow fields. (\textit{a},\textit{b}) smooth-wall reference case at $y^+= 5$, scaled with $u_\tau$ at $y^+ = 0$; (\textit{c},\textit{d}) slip-length simulation UWV6 at $y^+ = 5$, scaled with $u_\tau$ at $y^+ = 0$; (\textit{e},\textit{f}) the same snapshot as (\textit{c},\textit{d}), but now for the wall-parallel plane   $y^+ + \ell_T^+= 5$, scaled with $u_\tau$ at the origin for turbulence, $y^+ = - \ell_T^+$.}
\label{fig:comp_snaps}
\end{figure}

{The collapse of the mean velocity, r.m.s. fluctuations and Reynolds stress profile to the smooth-wall data for cases UVW1--UWV6,} shown in figures~\ref{fig:comp_uwv}(\textit{d}--\textit{f}), indicates that the near-wall turbulent dynamics remain smooth-wall-like, and that $\ell_T^+$ fully describes the effect of the virtual origins on the turbulence~\citep{Garcia-Mayoral2019}. It could, however, be argued that energy might be organized differently yet provide the same r.m.s. values. Figure~\ref{fig:comp_spectra} portrays the premultiplied energy spectra at $y^++\ell_T^+ = 15$ for several cases along with that of a smooth-wall flow at $y^+\approx 15$. For cases UWV1--UWV6, shown in figure~\ref{fig:comp_spectra}(\textit{e}--\textit{h}), the distribution of energy among different length scales is the same as in flows over a smooth wall, which supports the idea that the near-wall turbulence dynamics remain essentially smooth-wall-like. The same is true for cases V1, V2, UV1 and UV2, figure~\ref{fig:comp_spectra}(\textit{a}--\textit{d}), which {were discussed earlier in \S\,\ref{subsec:originForTurb}} and have $\ell_T^+ = 0$. Additionally, figure~\ref{fig:comp_snaps} compares snapshots of $u^{\prime+}$ and $v^{\prime+}$ for the flow over a smooth wall at $y^+ = 5$ with those for case UWV6 at two wall-parallel planes, $y^+ = 5$ and $y^+ +\ell_T^+ = 5$. The fluctuations at $y^+ = 5$ are portrayed in figures~\ref{fig:comp_snaps}(\textit{c},\textit{d}), and are scaled with $u_\tau$ measured at $y^+ = 0$. On the other hand, the fluctuations at $y^+ +\ell_T^+ = 5$, shown in figures~\ref{fig:comp_snaps}(\textit{e},\textit{f}), are scaled with $u_\tau$ measured at $y^+ = -\ell_T^+$. The figure demonstrates that there is no qualitative visual change in the flow when the snapshots from the smooth-wall flow are compared to those from the slip-length simulation at the equivalent height, i.e.\ comparing the smooth-wall flow at $y^+ = 5$ with case UWV6 at $y^+ + \ell_T^+ =5$, measuring $u_\tau$ accordingly. However, if the snapshots from the slip-length simulation are compared to the smooth-wall case at the same height above the reference plane $y^+=0$, using $u_\tau$ measured at $y^+=0$ in both cases as is often done in the literature, an apparent intensification of the fluctuations relative to the smooth-wall case can be observed, particularly for  $v^{\prime+}$, as shown in figures~\ref{fig:comp_snaps}(\textit{c},\textit{d}). This further supports the idea that the near-wall turbulence dynamics remain essentially smooth-wall like, except for the shift of origin $\ell_T^+$. Case UWV6 is used an example, because it has the deepest virtual origin for turbulence, $y^+\approx-4$, and the effect is more pronounced, but the same can also be observed for cases UWV1--UWV5.

\subsection{Separating the effect of the virtual origin experienced by the mean flow from that experienced by streamwise velocity fluctuations}\label{subsec:streamwiseOrigin}

\begin{figure}
\centering
\includegraphics{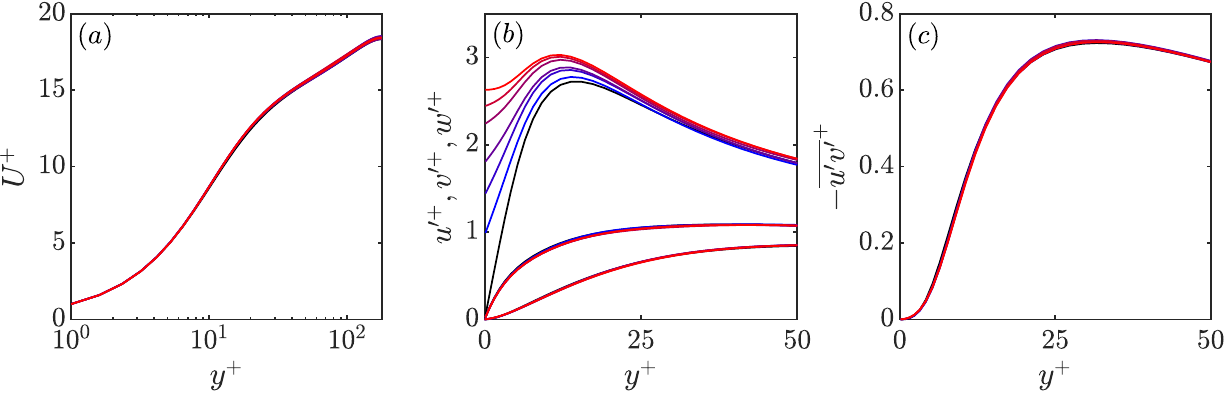}
\caption{Mean velocity profiles, r.m.s. velocity fluctuations and Reynolds shear stress profiles for simulations with slip on the streamwise fluctuations {but not on the mean flow, i.e. $\ell^+_{x,m} = 0$}. {Black, smooth-wall reference data; blue to red, cases UM1--UM6 with increasing slip on the streamwise fluctuations.}}
\label{fig:uprime_only}
\end{figure}

In \S\,\ref{subsec:originForTurb}, we observed for cases V1 and V2 that increasing $\ell_u^+$ resulted in an increase in the maximum value of $u^{\prime+}$ and a deviation of its profile away from the smooth-wall data. This leads to the question of why this is the case, to what extent the peak will continue to increase on increasing $\ell_u^+$, and what effect this has on the flow beyond simply modifying $u^{\prime+}$. More importantly, since this behaviour appears to occur independently of the mean flow, we also wish to address the question of whether or not the virtual origin perceived by the streamwise velocity fluctuations, and their apparent intensification, has any significant effect on $\Delta U^+$. To answer these questions, we carry out a series of simulations with no slip on the mean flow, i.e. $\ell_U^+=0$, and gradually increase the slip length applied to the streamwise fluctuations from $\ell_x^+=5$ to $\ell_x^+=\infty$, the latter being equivalent to a free-slip condition.  {These are simulations UM1--UM6. They} would highlight any effects caused by deepening the virtual origin experienced by the streamwise fluctuations. Results for these simulations are shown in figure~\ref{fig:uprime_only}. {Note that} the virtual origin experienced by the streamwise velocity fluctuations has no significant effect on the mean velocity profile, $v^{\prime+}$, $w^{\prime+}$ or the Reynolds stress profile, even for an infinite slip length. This demonstrates that the streamwise fluctuations play a negligible role in setting the origin for turbulence, i.e. $\ell_T^+ = 0$, and $\Delta U^+ = 0$. The peak value of $u^{\prime+}$ increases with the streamwise slip, but, even when $\ell_x^+=\infty$, {is not} much larger than the smooth-wall value. The $y$-location of this peak also does not change significantly. The gradual increase in the peak value is likely due to the greater $y$-range over which the streamwise fluctuations near the wall are brought to zero by viscosity. As $\ell_x^+$ is increased, the streamwise fluctuations experience a deeper virtual origin and have more room to decay to zero more slowly. This changes the slope of $u^{\prime+}$ near the wall and results in the gradual increase observed in the peak value, but has no other effect on the turbulence. This can, again, be confirmed from the premultiplied energy spectra for these cases, given  figure~\ref{fig:comp_spectra}(\textit{i}--\textit{l}). Except for case UM6, the simulation with infinite slip, the spectra for all cases matches very well to the smooth-wall reference data. There are some deviations in the contours for case UM6, but the peak location and overall distribution of energy among length scales still remains essentially smooth-wall-like. {It is perhaps surprising that} the Reynolds stress, $-\overline{u'v'}^+$, exhibits no change at all, given that $u^{\prime+}$ changes quite noticeably near the wall. However, since $v^{\prime+}$ is so small in the immediate vicinity of the wall, and the shape of its profile remains unmodified, the change in the Reynolds stress is negligible.

The above behaviour can be discussed in terms of the near-wall-cycle structures~\citep{Hamilton1995,Waleffe1997}. The quasi-streamwise vortices and streaks interact in a quasi-cyclic process in which the vortices act to sustain the streaks through sweeps and ejections of high- and low-speed fluid, respectively. In terms of the r.m.s. velocity fluctuations, the streaks are related to $u^{\prime+}$, while the quasi-streamwise vortices generate mainly $v^{\prime+}$ and $w^{\prime+}$. Since applying a slip length in the streamwise direction has no effect on the spanwise or wall-normal velocities, this means that the $y$-location of the quasi-streamwise vortices cannot change with respect to the domain boundary. If the vortices are unaffected, and the streaks are sustained by the vortices, there cannot, therefore, be a substantial change in the location or magnitude of the peak value of $u^{\prime+}$. This would explain why the origin for turbulence seems to be independent of the origin for the streamwise velocity fluctuations, at least in the regime where $\ell_u^+ \geq \ell_T^+$.

\begin{figure}
\centering
\includegraphics{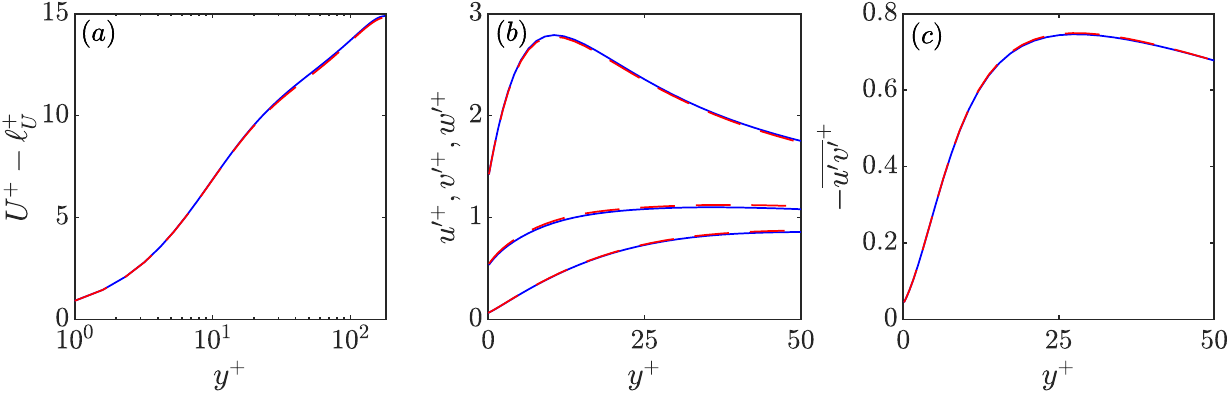}
\caption{Mean velocity profiles, r.m.s. velocity fluctuations and Reynolds shear stress profiles for simulations UWV6 (blue lines) and UWV6M (red dashed lines), which have the same slip lengths applied to the velocity fluctuations but different slip lengths applied to the mean flow.}
\label{fig:galilean}
\end{figure}

{So far, we have shown that applying a slip length to the streamwise flow appears to {have} an effect that is independent of the spanwise and wall-normal velocities. Further, applying a slip length to the streamwise fluctuations alone has no effect on the mean flow {or the turbulent fluctuations other than the effect on $u^{\prime+}$ that has no further consequence discussed above.} This suggests {that} the virtual origin experienced by the mean flow is independent of the virtual origin experienced by the streamwise fluctuations. The only streamwise origin relevant to $\Delta U^+$ would {therefore} be that experienced by the mean flow, and not the origin experienced by the near-wall streaks, and the streaks appear not to play a significant role in determining the drag.} We shall refer to this as the streaks being `inactive' with respect to the change in drag. {This is consistent with the idea proposed by \citet{Luchini1996} that turbulence as a whole has one origin, and the other important origin is the one for the mean flow, as we discussed in \S\,\ref{subsec:originForTurb}.} {We check this using simulations UWV6 and UWV6M, with the same set of slip-length coefficients for the fluctuating velocity components, but a different slip on the mean velocity.} {As shown in table~\ref{tab:slipLengthsVOs}, for the velocity fluctuations $(\ell_x^+,\ell_z^+,\ell_y^+)=(4.0,6.0,2.0)$ in both cases, but for the mean flow is $\ell_{x,m}^+=4.0$ and 10.0, respectively. The statistics for these simulations are portrayed in figure~\ref{fig:galilean}.} {The figure} shows that once $\ell_U^+$ is subtracted from the mean velocity, the mean velocity profile and the other statistics portrayed are identical for both simulations. This demonstrates that if the mean flow experiences a virtual origin different from the one of the streamwise fluctuations, this causes no change to the turbulence itself. This is because the additional mean velocity in case UWV6M corresponds simply to a Galilean shift of the flow compared to case UWV6. In combination with the fact that changing the virtual origin perceived by the streaks has no effect on the drag, this confirms that the important parameter in determining $\Delta U^+$ is $\ell_U^+$, and not $\ell_u^+$. {Note that} actual textures may not impose different virtual origins on the mean flow and the streamwise fluctuations, as is done in case UWV6M, but {its comparison with case UWV6} demonstrates which streamwise origin is physically relevant to $\Delta U^+$, and hence the drag. 

\subsection{Predicting the origin for turbulence from the virtual origins experienced by the three velocity components}

In the preceding discussion, we have shown that it is possible to displace the turbulence to its virtual origin at $y^+=-\ell_T^+$ by imposing virtual origins for the three velocity components. {We have demonstrated that the turbulence statistics and mean velocity profile essentially collapse to the smooth-wall data when rescaled by the friction velocity at $y^+=-\ell_T^+$ and measured from that same height. Since we have shown that $\ell_u^+$ has no effect on $\Delta U^+$, at least in the regime where $\ell_u^+ \geq \ell_T^+$, and that $\ell_U^+$ has no effect on the origin perceived by the turbulence, it follows that $\ell_T^+$ depends only on $\ell_v^+$ and $\ell_w^+$. {We have also observed} that the  resulting location of the origin for turbulence exhibits two distinct regimes, with respect to the virtual origins experienced by $v$ and $w$. The first is when the virtual origin for $v$ is deeper than or equal to the virtual origin for $w$, i.e.\ $\ell_v^+ \geq \ell_w^+$. With reference to table~\ref{tab:slipLengthsVOs}, if  $\ell_v^+ \geq \ell_w^+$, then $\ell_T^+\approx\ell_w^+$, which can be observed for cases V1, V2, UV1, UV2, UWV1, UWV2 and UWV4. The second regime occurs when the origin for $w$ is deeper than the origin for $v$, i.e. $0 < \ell_v^+< \ell_w^+$, for example, in cases UWV3, UWV5, UWV6 and UWV3H. We then find that turbulence perceives an origin intermediate between $\ell_v^+$ and $\ell_w^+$, such that  $ \ell_v^+< \ell_T^+<\ell_w^+$. This is in agreement with the physical arguments presented in \S\,\ref{sec:theory}. We now wish to infer an expression that can be used to predict the virtual origin for turbulence a priori from the virtual origins for $v$ and $w$. 

{Since we are interested in finding an expression for $\ell_T^+$ in terms of the virtual origins perceived by $v$ and $w$, and not the slip-length coefficients, it would be more appropriate to express the saturation in terms of $\ell_w^+$ rather than $\ell_z^+$. Following \citet{Fairhall2018}, but now taking into account the curvature of the $w^{\prime+}$ profile, we revisit the expression for the effective spanwise slip (\ref{eq:lzeff}), and arrive at the following empirical relation}
\begin{equation}
\ell_{w,eff}^+\approx \frac{\ell_w^+}{1+\ell_w^+/5},\label{eq:lweff}
\end{equation} 
{which is analogous to (\ref{eq:lzeff}) for $\ell_z^+$, but asymptotes to a value of 5 instead of 4.} Figure~\ref{fig:modified_saturation} is an alternative portrayal of the data from \citet{Busse2012} presented earlier in figure~\ref{fig:BS_saturation}(\textit{b}), but this time using $\ell_{w,eff}^+$ instead of $\ell_{z,eff}^+$ to calculate $\Delta U^+$. The figure shows excellent agreement between $\Delta U^+$ and the difference $\ell_x^+ - \ell_{w,eff}^+$, with the data for low $\Rey_{\tau,0}$ and high $\Delta U^+$ deviating again as discussed for figure~\ref{fig:BS_saturation}(\textit{b}). {The results show that when $\ell_v^+=0$, equation (\ref{eq:lweff}) gives an accurate prediction for the origin for turbulence, $\ell_{T,pred}^+$.} 

\begin{figure}
\centering
\vspace{2mm}\includegraphics{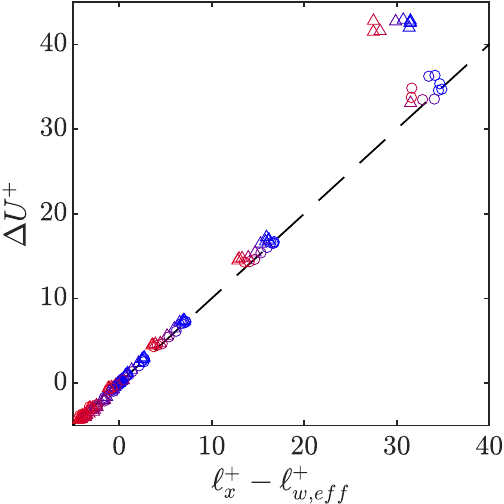}
\caption{Alternative portrayal of the data from \citet{Busse2012} presented in figure~\ref{fig:BS_saturation}(\textit{b}), with $\Delta U^+$ now a function of $\ell_x^+ - \ell_{w,eff}^+$. Triangles, simulations at $\Rey_{\tau,0} = 180$; circles, simulations at $\Rey_{\tau,0}  = 360$. From blue to red, increasing $\ell_{w,eff}^+$. The dashed line represents $\Delta U^+ = \ell_x^+ - \ell_{w,eff}^+$.}
\label{fig:modified_saturation}
\end{figure} 

\begin{figure}
\centering
\includegraphics[width=0.8\textwidth]{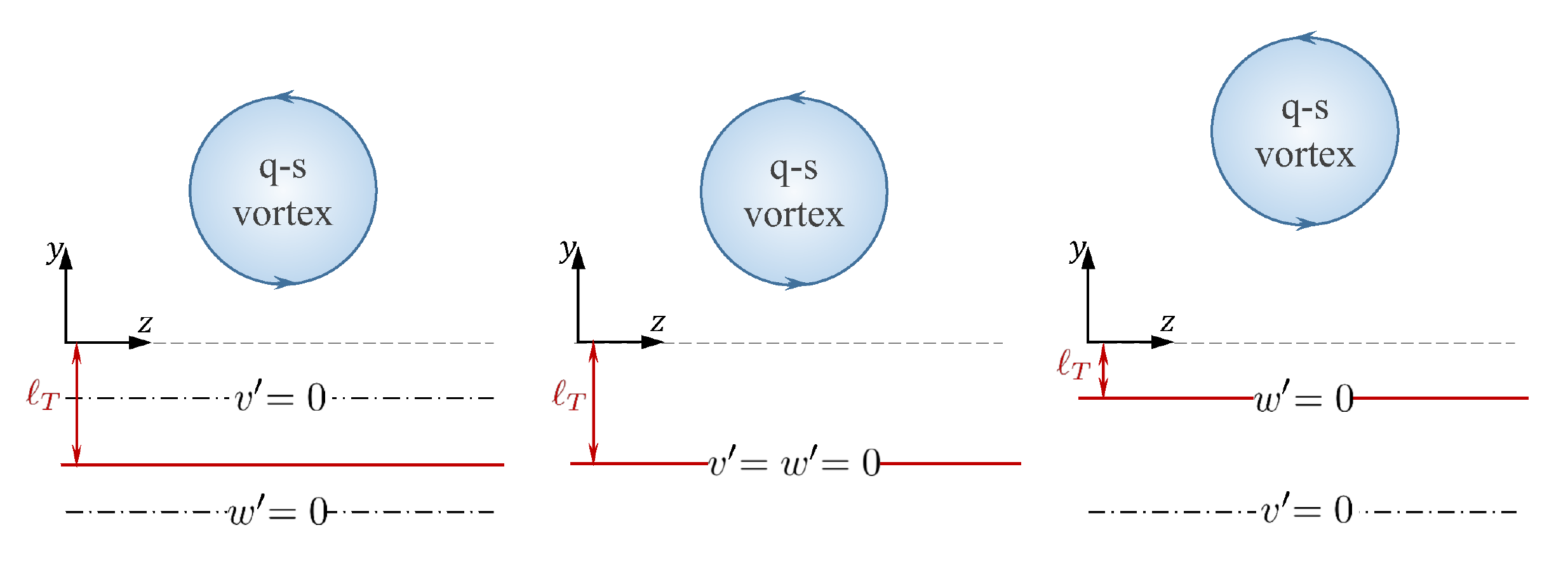}
\mylab{-10.7cm}{3cm}{($a$)}%
\mylab{-7cm}{3cm}{($b$)}%
\mylab{-3.8cm}{3cm}{($c$)}%
\caption{Schematics of the location of the origin for turbulence, $y^+ = -\ell_T^+$, when imposing different origins for the spanwise and wall-normal velocities. The planes where $v^\prime=0$ and $w^\prime=0$ correspond to the imposed virtual origins, $y^+ = - \ell_v^+$ and $y^+=-\ell_w^+$, respectively  The origin for turbulence, $y^+ = -\ell_T^+$, is represented by the red line. ($a$) $\ell_v^+<\ell_w^+$, ($b$) $\ell_v^+ = \ell_w^+$, ($c$) $\ell_v^+>\ell_w^+$. Note that in each case, the distance between the centre of the quasi-streamwise vortices (q-s vortex) and the plane $y^+=-\ell_T^+$ is the same.}
\label{fig:VO_final}
\end{figure}

However,  $\ell_{T,pred}^+$ should also depend on $\ell_v^+$, because the saturation in the effect of $\ell_w^+$ is a direct result of the plane at which $v$ perceives an impermeable wall. {Therefore, {if} the saturation of the effect of $\ell_w^+$  is evaluated with respect to the plane at which $v$ appears to vanish, i.e.\ $y^+=-\ell_v^+$, rather than $y^+=0$, it should be possible to predict the virtual origin for turbulence in the {regime $0<\ell_v^+<\ell_w^+$}~\citep{Gomez-de-Segura2018a}.} This regime is sketched in figure~\ref{fig:VO_final}(\textit{a}), where the virtual origin for turbulence lies between the virtual origins for $v$ and $w$, with $\ell_v^+ < \ell_{T,pred}^+ < \ell_w^+$. Then, $\ell_{T,pred}^+$ would be given by
\begin{equation}
\ell_{T,pred}^+ \approx \ell_v^+ + \frac{(\ell_w^+ - \ell_v^+)}{1+(\ell_w^+ - \ell_v^+)/5}.\label{eq:lT_lv}
\end{equation}
On the other hand, as demonstrated by \citet{Gomez-de-Segura2018a}, if $\ell_v^+\approx\ell_w^+$ then no saturation in the effect of $\ell_w^+$ occurs and $\ell_{T,pred}^+\approx\ell_w^+$, as sketched in figure~\ref{fig:VO_final}(\textit{b}). If we increase $\ell_v^+$ further, such that $\ell_v^+>\ell_w^+$, we would not expect the quasi-streamwise vortices to approach the surface further, since their first-order effect is to induce a spanwise velocity at the reference plane. Even if $v$ was allowed to penetrate freely through the reference plane, the quasi-streamwise vortices would require some amount of spanwise slip in the first place to approach this plane. Therefore, when $\ell_v^+\geq\ell_w^+$, we would expect $\ell_T^+\approx\ell_w^+$. This is confirmed in our simulations UWV1, UWV2, and UWV4. This regime is sketched in figure~\ref{fig:VO_final}(\textit{c}).  Combining (\ref{eq:lT_lv}) and the preceding argument for the regime where $\ell_v^+\geq\ell_w^+$, a general expression for approximating $\ell_T^+$ from $\ell_w^+$ and $\ell_v^+$ would be
\begin{equation}
\ell_{T,pred}^+ \approx
    \begin{dcases*}
      \ell_v^+ + \frac{(\ell_w^+ - \ell_v^+)}{1+(\ell_w^+ - \ell_v^+)/5} & if $\ell_w^+>\ell_v^+$,\\
     \ell_w^+ & if $\ell_w^+\leq \ell_v^+$.\label{eq:lTp}
    \end{dcases*}
\end{equation}
When $\ell_{T,pred}^+$ is predicted from the values of $\ell_v^+$ and $\ell_w^+$, which are known a priori from $\ell_y^+$ and $\ell_z^+$, it shows excellent agreement with the value of $\ell_T^+$ measured a posteriori for the cases presented thus far, as shown in table~\ref{tab:slipLengthsVOs}. In all these cases, $\Delta U^+$ is given by the linear law (\ref{eq:DUlT}), that is, the difference $\ell_U^+-\ell_T^+$.

{A key point epitomised by (\ref{eq:lTp}) is that the only relevant parameters are the relative positions of the virtual origins of $u$ and $w$ relative to the plane where $v$ appears to vanish. The classical understanding, as first proposed by~\citet{Luchini1991}, is that the only relevant parameter is the difference between streamwise and spanwise protrusion heights. However, the results of \citet{Busse2012} show that this is not the case. We argue that the plane where $v$ appears to vanish (or alternatively, how much the flow can transpire through the reference plane from which the tangential virtual origins are measured) is also important. The result is an extension of Luchini's theory where, rather than on the difference between the virtual origins perceived by the tangential velocities, $\Delta U^+$ depends on their positions relative to that perceived by the wall-normal velocity, regardless of the plane taken as reference.}

\subsection{Scaling with Reynolds number {and domain size}}\label{subsec:reEffects}

As discussed in \S\,\ref{sec:theory}, the universal parameter for quantifying the performance of drag-reducing surfaces is $\Delta U^+$. The idea is that so long as the texture size of a given surface, $L^+$, is fixed in wall units, then so would be $\ell_u^+$, $\ell_v^+$ and $\ell_w^+$, and $\Delta U^+$ should remain essentially independent of $\Rey_\tau$. {In our simulations, we impose different virtual origins on each velocity component, $\ell_u^+$, $\ell_v^+$ and $\ell_w^+$, and so we wish to determine whether this effect scales in wall units {for varying Reynolds numbers.} To verify this, we {conduct} two simulations at different Reynolds numbers, $\Rey_\tau = 180$ and 550, but {keep} the virtual origins constant in wall units, see cases UWV3 and UWV3H in table~\ref{tab:slipLengthsVOs}.} For the two cases considered here, the slip lengths $\ell_x^+$, $\ell_y^+$ and $\ell_z^+$ are identical for both Reynolds numbers. Note that, in general, this will not necessarily be the case, since the one-to-one a priori relationship discussed in {\S\,\ref{subsec:vo_setup}} between the Robin slip-length coefficients in equation  (\ref{eq:slip_lengths}), $\ell_x^+, \ell_y^+$ and $\ell_z^+$, and the resulting virtual origins, $\ell_u^+, \ell_v^+$ and $\ell_w^+$, will slightly change with the Reynolds number. However, these differences are consistent with the change in the turbulence statistics over a smooth wall as a result of varying the Reynolds number~\citep[see e.g.][]{Moser1999}, as shown figure~\ref{fig:comp180550}. After shifting the mean velocity profile and r.m.s. velocity fluctuations by $\ell_T^+$ and rescaling them by the friction velocity at $y^+=-\ell_T^+$, they essentially collapse to the smooth-wall data. For both Reynolds numbers, the value of $\Delta U^+ = \ell_U^+ - \ell_T^+$ measured a posteriori is 1.3 (see table~\ref{tab:slipLengthsVOs}), indicating that $\Delta U^+$ is indeed independent of the Reynolds number for fixed values of the virtual origins $\ell_u^+$, $\ell_v^+$ and $\ell_w^+$ in wall units. However, the measured drag reduction,  $\mbox{\textit{DR}}$, varies between the two cases, as expected. From (\ref{eq:DR2}), $\mbox{\textit{DR}}$ is smaller at higher $\Rey_\tau$, due to the increase in $U_{\delta_0}^+$ with $\Rey_\tau$, even though we observe no change in $\Delta U^+$. 

\begin{figure}
\centering
\includegraphics{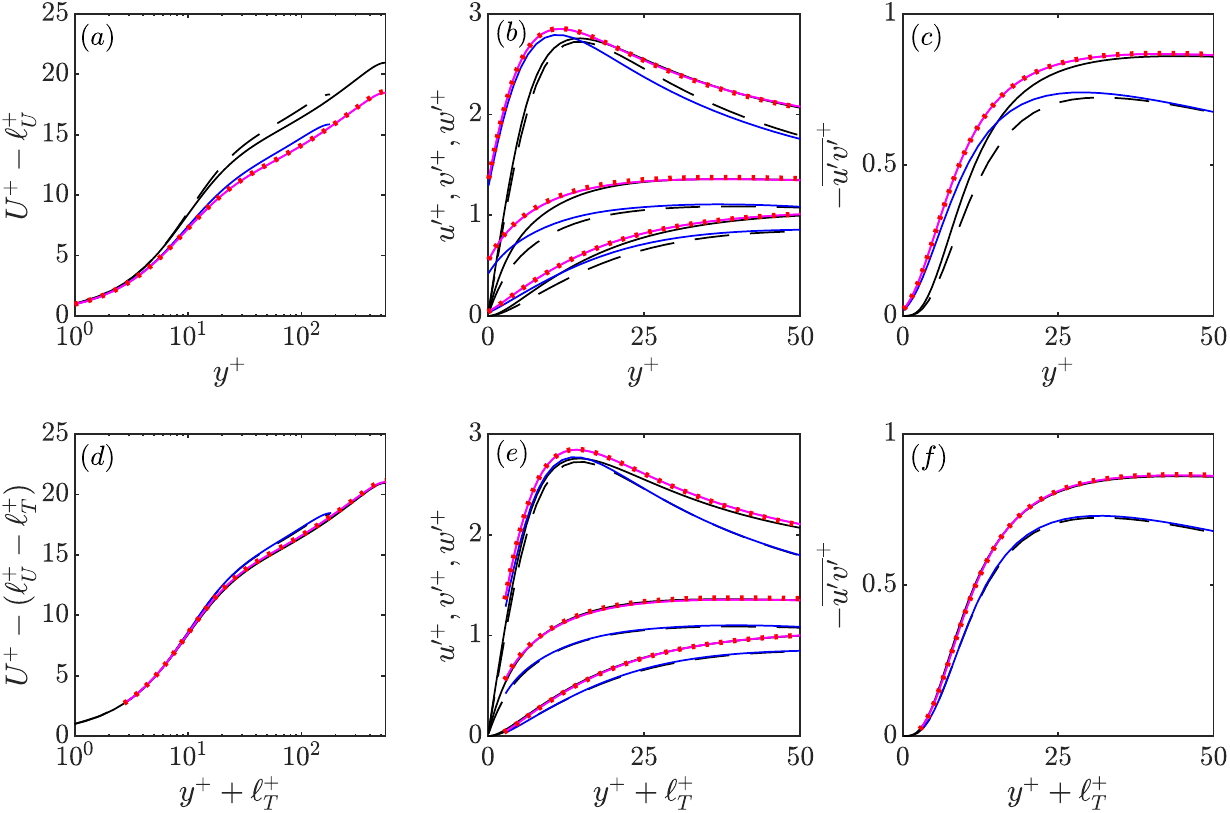}
\caption{{Mean velocity profiles, r.m.s. velocity fluctuations and Reynolds shear stress profiles for slip-length simulations UWV3, UWV3H and UWV3HD. These simulations have the same virtual origins, in wall units, for each velocity component, $(\ell_u^+,\ell_w^+,\ell_v^+) = (3.6,2.9,1.9)$, { but} UWV3 is conducted at $\Rey_\tau \simeq 180$, whereas UWV3H and UWV3HD are conducted at $\Rey_\tau \simeq 550$. {UWV3HD} has a larger domain size in the wall-parallel directions, {$8\pi\times3\pi$ instead of $2\pi\times\pi$}. (\textit{a}--\textit{c}), scaled with the friction velocity at the reference plane, $y^+=0$; (\textit{d}--\textit{f}) shifted in $y^+$ by $\ell_T^+$ and scaled with the friction velocity at the origin for turbulence, $y^+=-\ell_T^+$.} {Smooth-wall reference data is portrayed at ({\color{black}\dashed}) $\Rey_\tau \simeq 180$ and ({\color{black}\full}) $\Rey_\tau \simeq 550$; {\color{blue}\full}, case UWV3; {\color{matlabmagenta}\full}, case UWV3H; {\color{red}\dotted}, case UWV3HD.}}
\label{fig:comp180550}
\end{figure}

{Our simulations domains, $2\pi \times \pi$, are sufficiently large to capture the key turbulence processes and length scales of the near-wall and log-law regions of the flow~\citet{Lozano-Duran2014}, but scales larger than this will be unresolved. To verify that virtual origins interact only with the smaller scales that reside near the wall, we conduct an additional simulation at $\Rey_\tau \simeq 550$, UWV3HD, with the same parameters as UWV3H but a domain size $8\pi \times 3\pi$. The results shown in figure~\ref{fig:comp180550} are indistinguishable, suggesting that the origin-offset mechanism does not interact with the larger, outer turbulence scales, other than by the shift in origin.}

\subsection{Departure from smooth-wall-like turbulence}\label{subsec:breakdown}

The fundamental idea behind the proposed virtual-origin framework, as discussed in {\S\,\ref{subsec:vo_setup}}, is that when we impose virtual origins on each velocity component, we assume that the shape of each r.m.s.\ velocity profile remains smooth-wall-like independently of the others. For this assumption to hold, the near-wall turbulence cycle should be left essentially unaltered. Otherwise, these profiles will no longer be smooth-wall-like. {As a guide, we can say that the virtual origins should be smaller than the smallest eddies of near-wall turbulence. As discussed in \S\,\ref{sec:introduction}, this would be the quasi-streamwise vortices, with  diameter and distance to the surface both of order 15 wall units~\citep{Robinson1991,Schoppa2002}}. This would then serve as a rough limit for the applicability of this framework. The cases presented thus far have all been within this regime, and we have demonstrated that the flow remained essentially smooth-wall-like, once the virtual origin for turbulence, $\ell_T^+$, was accounted for. It was also possible to predict $\ell_T^+$ from the virtual origins a priori.

\begin{figure}
\centering
\includegraphics{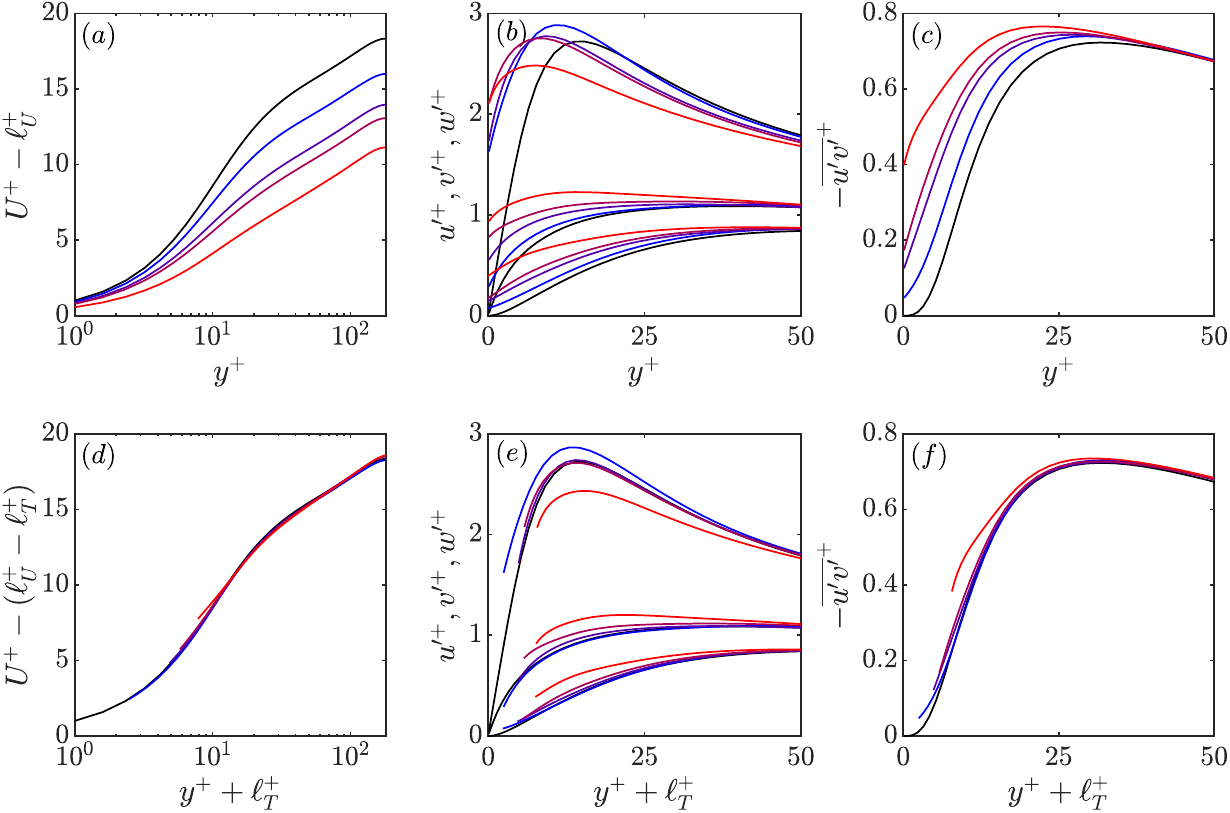}
\caption{Mean velocity profiles, r.m.s. velocity fluctuations and Reynolds shear stress profiles for simulations with slip-length boundary conditions applied to all three velocity components. Here, the values of the slip-length coefficients are relatively large, e.g.\ up to $\ell_x^+,\ell_y^+,\ell_z^+ \approx 10$. (\textit{a}--\textit{c}) scaled with the friction velocity at the reference plane, $y^+=0$; (\textit{d}--\textit{f}) shifted in $y^+$ by $\ell_T^+$ and scaled with the friction velocity at $y^+=-\ell_T^+$. Black lines, smooth-wall reference data; blue to red lines, cases UWVL1--UWVL4.}
\label{fig:breakdown}
\end{figure}

{To better understand} the limits of this framework, we conduct a series of simulations where the imposed virtual origins are relatively large, e.g.\ up to $\ell_u^+,\ell_w^+ \approx 6$ and $\ell_v^+ \approx 11$. These are cases UWVL1--UWVL4, and their mean velocity profiles, r.m.s. velocity fluctuations and Reynolds stress profiles are shown in figure~\ref{fig:breakdown}. We see that when the imposed virtual origins become too large, the r.m.s. velocity fluctuations and Reynolds stress profile no longer remain smooth-wall-like. In these cases, as we increase the depth of the virtual origins, specifically for $v$ and $w$, the quasi-streamwise vortices approach the reference plane $y^+=0$ to such an extent that they are, in fact, ingested by the domain boundary. The whole near-wall cycle is then fundamentally disrupted, changing the nature of the flow near the wall. This is most apparent for cases UWVL3 and UWVL4. The premultiplied energy spectra for these cases, shown in figure~\ref{fig:comp_spectra_breakdown}(\textit{a--d}), indicate that there can be a dramatic change in the distribution of energy among length scales, compared to the smooth-wall case, when the imposed virtual origins are large. For example, in case UWVL4 there is a significant redistribution of energy  in the wall-normal velocity to larger spanwise and shorter streamwise wavelengths. This also occurs when the transpiration triggers the appearance of Kelvin--Helmholtz-like spanwise rollers~\citep[see e.g.][]{Garcia-Mayoral2011b,Gomez-de-Segura2019} or in the presence of roughness large enough to disrupt the near-wall cycle~\citep{Abderrahaman-Elena2019}. This increased spanwise coherence of $v^{\prime+}$ can also be observed in the snapshots of case UWVL4, which are compared to those of the smooth-wall reference case in figure~\ref{fig:snaps_KH}. A similar behaviour was also observed by \citet{Gomez-de-Segura2018a} when using a Stokes-flow model for the virtual layer of flow below $y^+=0$, rather than the Robin slip-length boundary conditions used here. 

\begin{figure}
\centering
\includegraphics{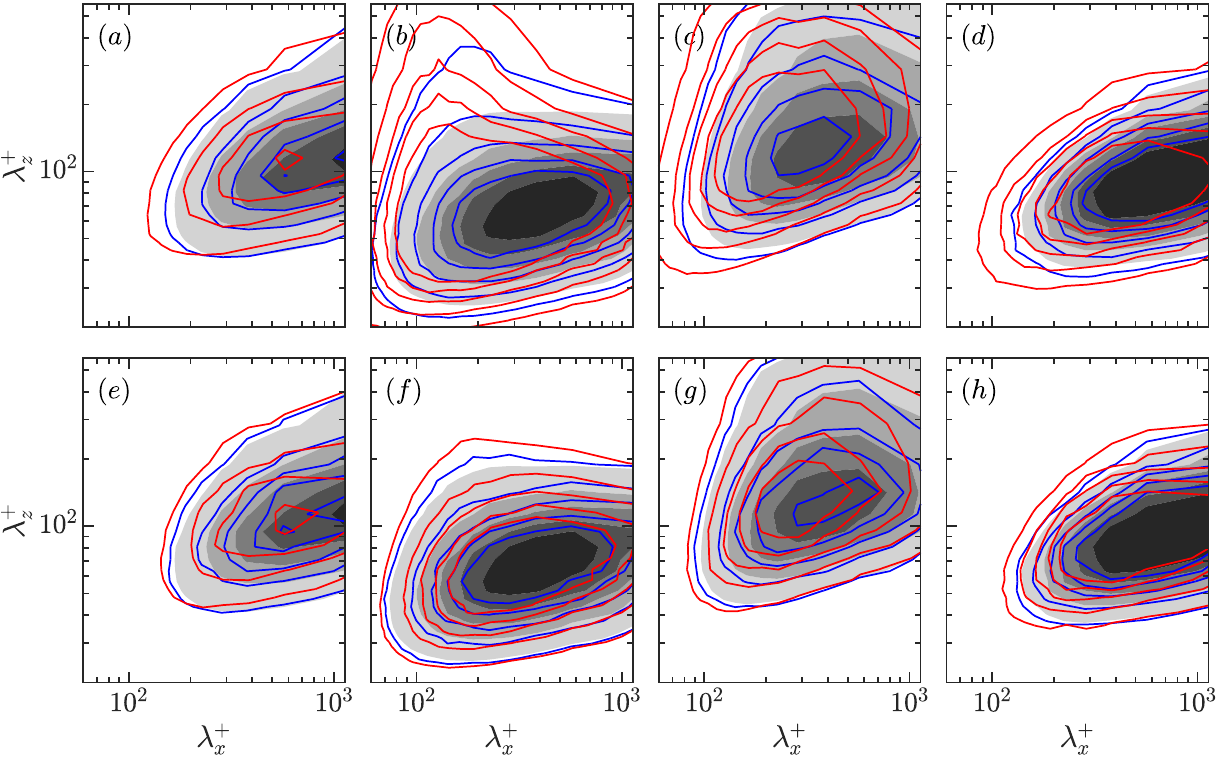}
\caption{Premultiplied two-dimensional spectral densities of $u^2$, $v^2$, $w^2$ and $uv$ at $y^+ + \ell_T^+ = 15$, normalised by $u_\tau$ at $y^+ = -\ell_T^+$, for various slip-length simulations (line contours), compared to smooth-wall data (filled contours) at $y^+ = 15$.  The shift $\ell_T^+$ is given in table~\ref{tab:slipLengthsVOs} for each case. (\textit{a}--\textit{d}) cases UWVL1 and UWVL4, with line colours as in figure~\ref{fig:breakdown}.  (\textit{e}--\textit{h}) cases WV1 and WV3, with line colours as in figure~\ref{fig:dragIncrease}. First column, $(k_x k_z E_{uu})^+$; second column, $(k_x k_z E_{vv})^+$; third column, $(k_x k_z E_{ww})^+$; fourth column, $-(k_x k_z E_{uv})^+$. The contour increments for each column are 0.3224, 0.0084, 0.0385 and 0.0241, respectively.}
\label{fig:comp_spectra_breakdown}
\end{figure}

\begin{figure}
\centering
\includegraphics{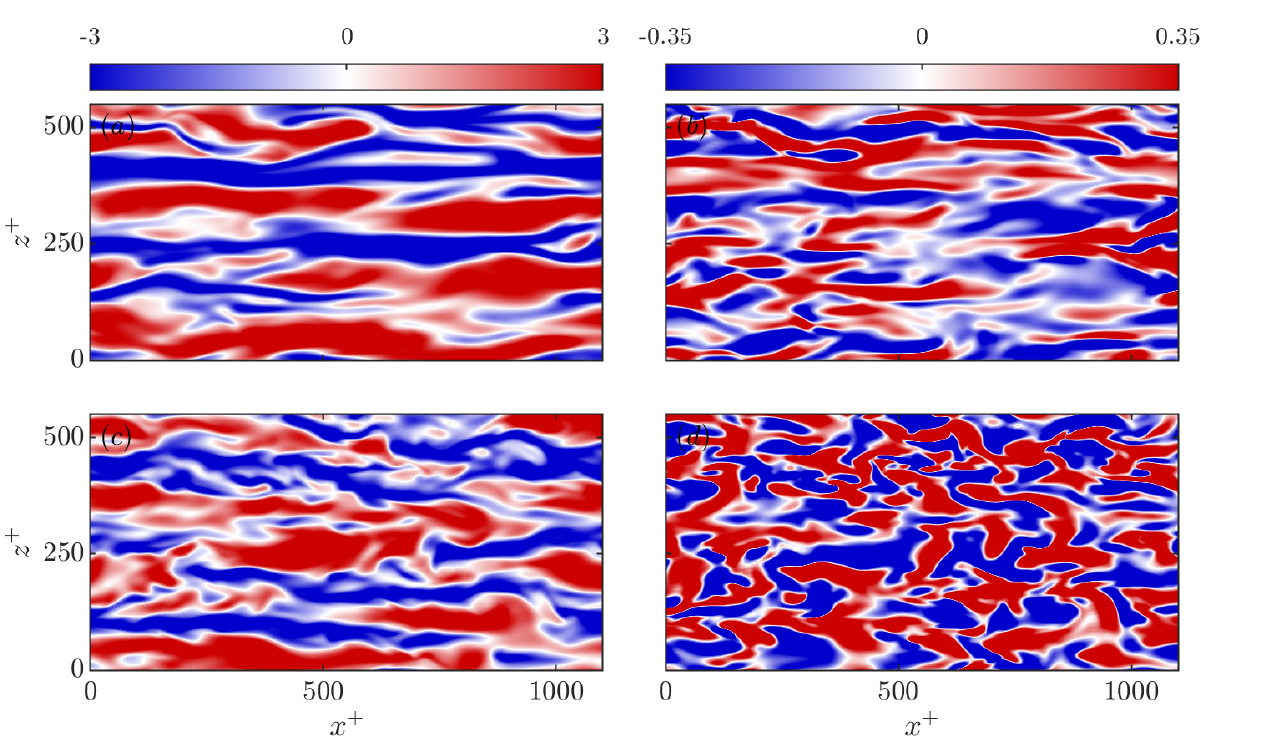}
\caption{Streamwise (\textit{a},\textit{c}) and wall-normal (\textit{b},\textit{d}) instantaneous velocity fluctuation flow fields. (\textit{a},\textit{b}) smooth-wall reference case at $y^+= 15$, scaled with $u_\tau$ at $y^+ = 0$; (\textit{c},\textit{d}) slip-length simulation UWVL4 at $y^+ +\ell_T^+ = 15$, scaled with $u_\tau$ at $y^+ = -\ell_T^+$.}
\label{fig:snaps_KH}
\end{figure}

{The} results of cases UWVL1--UWVL4, {suggest} that the virtual-origin framework holds {only for} $\ell_T^+ \lesssim 5$. {Beyond this point, the Reynolds stress profiles presented in figures~\ref{fig:breakdown}(\textit{c},\textit{f}) indicate that the near-wall turbulence is no-longer smooth-wall-like, and the underlying assumptions of the framework are no longer valid.} As discussed above, the origin for turbulence, $\ell_T^+$, should depend only on $\ell_v^+$ and $\ell_w^+$. Note, however, that it is more difficult to impose limits on $\ell_w^+$ and $\ell_v^+$ independently, because both spanwise slip and transpiration are required to increase $\ell_T^+$ beyond 5 wall units, as encapsulated by (\ref{eq:lTp}). For very large spanwise slip lengths without transpiration~\citep[e.g.][]{Busse2012}, the virtual-origin framework still holds, {and a saturation in the effect of $\ell_z^+$ is observed,} as discussed in \S\,\ref{sec:theory}. In turn, as we have seen in  cases V1, V2, UV1 and UV2, increasing $\ell_v^+$ beyond $\ell_w^+$ bears no consequence on $\ell_T^+$, no matter how large $\ell_v^+$.


\begin{figure}
\centering
\includegraphics{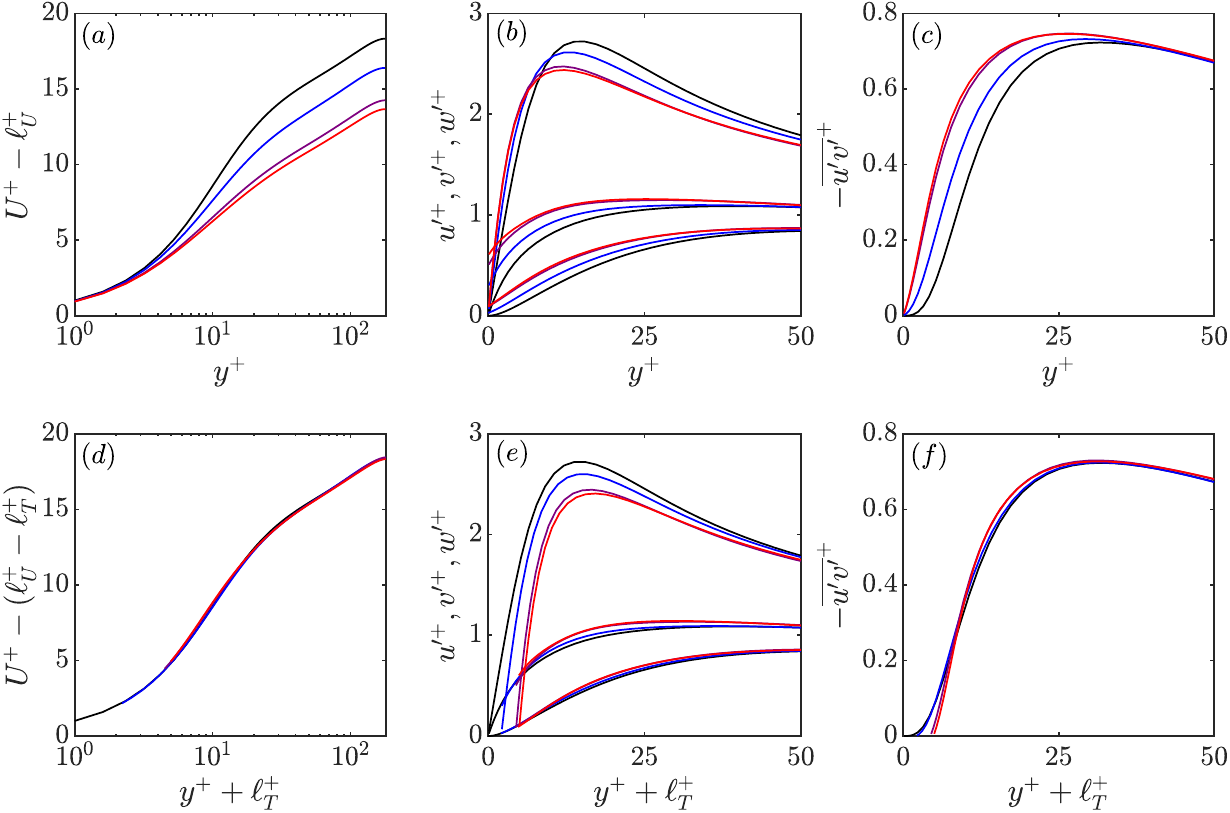}
\caption{Mean velocity profiles, r.m.s. velocity fluctuations and Reynolds shear stress profiles for simulations with slip-length boundary conditions applied to the spanwise and wall-normal velocity components only. (\textit{a}--\textit{c}) scaled with the friction velocity at the reference plane, $y^+=0$; (\textit{d}--\textit{f}) shifted in $y^+$ by $\ell_T^+$ and scaled with the friction velocity at $y^+=-\ell_T^+$. Black lines, smooth-wall reference data; blue to red lines, cases WV1--WV3.}
\label{fig:dragIncrease}
\end{figure}

The cases considered so far {satisfy} $\ell_u^+ \gtrsim \ell_T^+$, i.e.\ the streaks {perceive} a virtual origin at least as deep as {that} perceived by the quasi-streamwise vortices. In this regime, as discussed \S\,\ref{subsec:streamwiseOrigin}, the streamwise fluctuations have a greater $y$-range in which they are brought to zero by viscosity, as shown in figure~\ref{fig:uprime_only}, but otherwise the quasi-streamwise vortices and the turbulence remain smooth-wall like. {There is} sufficient room for the near-wall-cycle structures to reside, and no change in the turbulence dynamics is observed. In contrast, we now consider the opposite regime, where $\ell_u^+ < \ell_T^+$. {This would arguably be the case of interest for roughness, and has been shown to be the case when roughness is sufficiently small~\citep{Abderrahaman-Elena2019}.} We fix the origin for the streamwise velocity at $y^+=0$, i.e.\ $\ell_u^+ = 0$, and progressively increase the depth of the origin for turbulence below this plane, i.e.\ $\ell_T^+>0$. As portrayed in figure~\ref{fig:dragIncrease}, for cases WV1--WV3, we then  observe a gradual departure from smooth-wall-like turbulence. Note that $\ell_U^+<\ell_T^+$ in these cases, and so $\Delta U^+<0$, which would correspond to an increase in drag. Case WV1, with $\ell_T^+\approx 2$, appears to be the limiting case, in which  turbulence still remains essentially smooth-wall-like, as can be observed in figure~\ref{fig:dragIncrease}(\textit{d--f}),  and $\ell_{T,pred}^+$, calculated from equation (\ref{eq:lTp}), still provides a reasonable estimate for the origin for turbulence, as shown in table~\ref{tab:slipLengthsVOs}. However, increasing $\ell_T^+$ further results in clear differences between the r.m.s. velocity fluctuations and Reynolds stress profiles compared to the smooth-wall data. This can also be seen in the premultiplied energy spectra shown in figure~\ref{fig:comp_spectra_breakdown}(\textit{e--h}), where the distribution of energy among length scales is no longer smooth-wall-like. For example, the spectra of case WV3 indicates that the energy in the streamwise and spanwise velocity components is now shifted, on average, to shorter streamwise wavelengths. Schematically, the quasi-streamwise vortices would approach the reference plane, but the streaks would be constrained by the condition that $u^{\prime +}=0$ at $y^+=0$. The streaks, which are sustained by the vortices, would then no longer have sufficient room to reside above $y^+=0$, compared to the flow over a smooth wall, and would become squashed in $y$ and weakened. This, in turn, would restrict the whole near-wall turbulence dynamics, and cause the flow not to remain smooth-wall-like. From our simulations, this breakdown appears to occur when the virtual origin for turbulence is more than approximately 2 wall units deeper than the origin perceived by the streaks. Therefore, an additional constraint on the present virtual-origin framework would be that the imposed virtual origins should satisfy $\ell_T^+ \lesssim \ell_u^+ + 2$. {This is in agreement with the observation in \citet{Abderrahaman-Elena2019} that a virtual-origin framework alone cannot capture the effect of roughness on the flow once the roughness size is large enough that $\Delta U^+ \simeq -2$. Further, it highlights the limitations of  modelling the effects of drag-increasing surfaces, such as roughness, with virtual origins alone.} 



The breakdown of the virtual-origin framework and the subsequent departure from smooth-wall-like turbulence require further discussion. The homogeneous slip-length boundary conditions (\ref{eq:slip_lengths}) are an approximation of the apparent boundary conditions that real textured surfaces impose on the flow. They are a reasonable model so long as the characteristic texture size is small compared to the length scales of the turbulent eddies in the flow~\citep{Garcia-Mayoral2019}. As the texture size, $L^+$, is increased, we expect the apparent virtual origins that a given surface imposes on the flow to become deeper. However, on increasing $L^+$ further, flows over real textured surfaces eventually exhibit additional dynamical mechanisms, typically drag-degrading, such that the effect of the texture can no longer be approximated by a simple virtual-origin model. For example, as $L^+$ increases for superhydrophobic surfaces, the flow begins to perceive the texture as discrete elements, as opposed to a homogenised effect~\citep{Seo2016,Fairhall2019}, and the entrapped gas pockets can also be lost~\citep{Seo2018}, both of which fundamentally change the apparent boundary conditions imposed by the surface on the flow. For riblets, in turn, increasing the texture size can trigger the onset of Kelvin--Helmholtz-like rollers, which can have a strong drag-increasing effect on the flow~\citep{Garcia-Mayoral2011b}}. Importantly, the depth of the virtual origin for turbulence at which the present framework breaks down, i.e.\ $\ell_T^+ \approx 5$, could imply a texture size that would place a corresponding real surface in a regime beyond the onset of the failure mechanisms just mentioned. For instance, the onset of the Kelvin--Helmholtz-like instability in riblets can occur for $\ell_T^+ \gtrsim 1$~\citep{Garcia-Mayoral2011b}. In this case, the limits imposed by the operating window of the real surface are the most restrictive, and not the theoretical limits of the virtual-origin framework. Therefore, if the goal of a given simulation is to use virtual origins to model the effect on the flow of a real surface, it is crucial to keep in mind the texture size, and hence the magnitude of the virtual origins, at which the flow no longer perceives the surface in a homogenised fashion. This could, in many instances, be the actual limit up to which the virtual-origin framework can be feasibly applied.
 


\section{Active opposition control interpreted in a virtual-origin framework}\label{sec:opposition}

\begin{figure}
\centering
\includegraphics{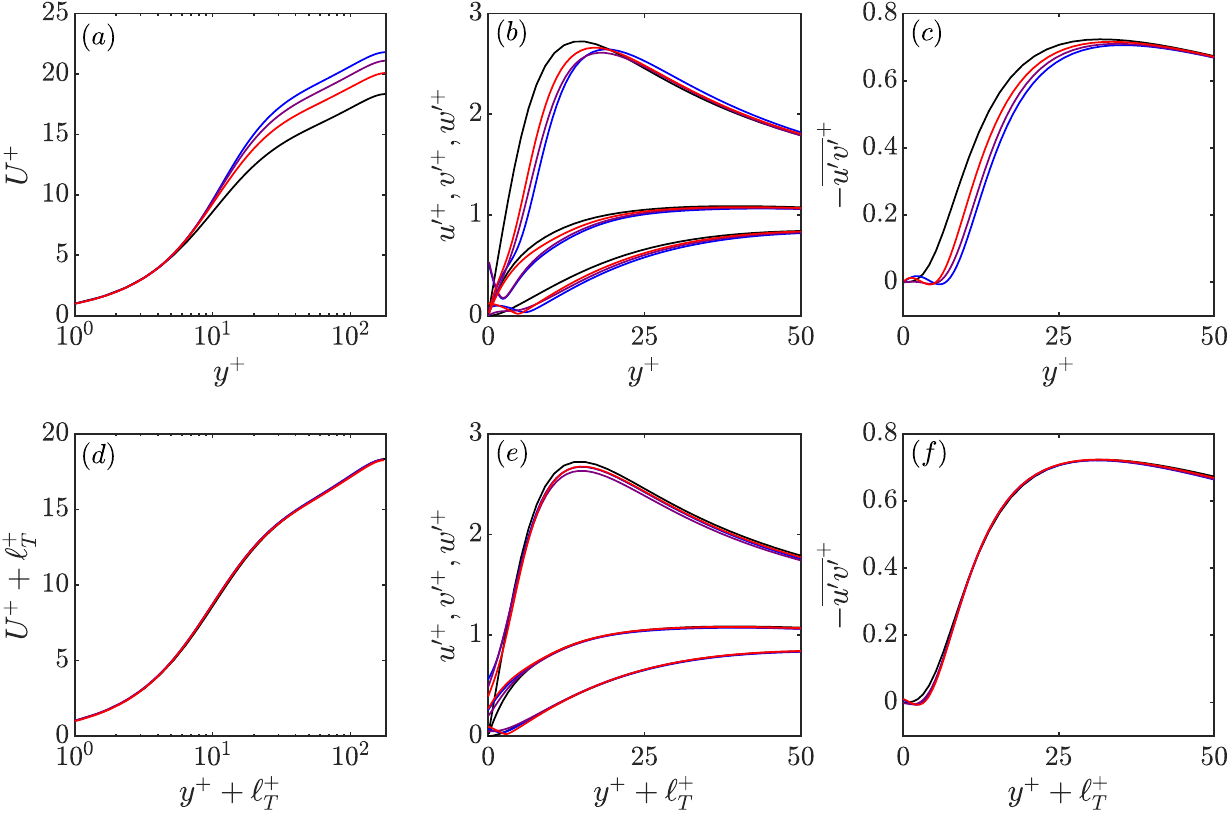}
\caption{Mean velocity profiles, r.m.s. velocity fluctuations and Reynolds shear stress profiles for simulations with opposition control on $w$ and $v$ (in various combinations), with the sensing plane at $y_d^+ = 7.8$. (\textit{a}--\textit{c}) scaled with the friction velocity at the reference plane, $y^+=0$; (\textit{d}--\textit{f}) shifted in $y^+$ by $\ell_T^+$ and scaled with the friction velocity at the origin for turbulence, $y^+=-\ell_T^+$. Note that in these cases, $\ell_T^+<0$ and therefore the origin for turbulence is above the plane $y^+=0$. Black lines, smooth-wall reference data; blue to red lines, $w$-$v$ control, $w$ control and $v$ control.}
\label{fig:opposition}
\end{figure}

{As we mentioned in \S\,\ref{sec:theory}, the findings of the original study on opposition control by \citet{Choi1994} suggest that the effect of the control was to cause an outward shift of the origin for turbulence with respect to the mean flow. This is precisely the idea behind the present virtual-origin framework, captured by (\ref{eq:DUlT}). We now assess if we can also explain the effect of opposition control on the flow with a virtual-origin framework.}

{We conduct three opposition control simulations, controlling $v$ alone, $w$ alone, and both $v$ and $w$, with the detection plane in each case at $y_d^+ = 7.8$, as outlined in \S\,\ref{subsec:oc_setup} and table~\ref{tab:opposition}.}  Raw results of these simulations are shown in figures~\ref{fig:opposition}(\textit{a--c}), where, as expected, we observe an outward shift in the turbulence statistics away from the domain boundary~\citep{Choi1994}.  We then measure the virtual origin for turbulence, $\ell_T^+$, a posteriori, using the method outlined in \S\,\ref{subsec:originForTurb}, and rescale the data with respect to the friction velocity at that origin, from equation (\ref{eq:utau}). {The results} shifted in $y^+$ by $\ell_T^+$ are included in figures~\ref{fig:opposition}(\textit{d--f}). Note that, for these cases, the origin for the mean flow is at the reference plane $y^+=0$, i.e.\ $\ell_U^+ = 0$, and so $\Delta U^+ = -\ell_T^+$, from (\ref{eq:DUlT}). The excellent collapse of the turbulence statistics in figure~\ref{fig:opposition}(\textit{d--f}) indicates that the effect of this active control technique is to cause an outward shift of the origin for turbulence away from the origin for the mean flow, and that turbulence does, indeed, remain smooth-wall-like except for this shift of origin. This suggests that it might be possible to consolidate the effect on the flow of other a wide variety of passive textures and active-control techniques in terms of a relative displacement of the virtual origin for turbulence and the virtual origin perceived by the mean flow, with the turbulence remaining otherwise smooth-wall-like.


While we have shown that for opposition control  it is possible to find a shift in the origin for turbulence, $\ell_T^+$, that results in a collapse of the turbulence statistics to the smooth-wall profiles, we now wish to see if it is possible to predict $\ell_T^+$ (and $\Delta U^+$) from the virtual origins perceived by the three velocity components, as we did for the slip-length simulations in \S\,\ref{sec:results}. {First we} establish where each velocity component would notionally perceive a virtual origin when the control is applied. Based on the discussion {in \S\,\ref{sec:theory}}, we assume that this would be the plane $y^+ = y_d^+/2$ for controlled velocity components, and $y^+=0$ for uncontrolled ones. In our slip-length simulations, in contrast, the apparent virtual origins were always at or below the domain boundary, e.g.\ at $y^+ = - \ell_w^+$, where $\ell_w^+\geq 0$. We retain the same nomenclature here, but now the sign of the virtual origins would be reversed, e.g.\  $\ell_w^+\leq 0$. For instance, in the case of $w$-$v$ control, the virtual origin perceived by the mean flow would be the domain boundary $y^+=0$, while the virtual origins perceived by $v$ and $w$ would be the plane $y^+ = 3.9$, yielding $\ell_U^+ = 0$ and $\ell_v^+=\ell_w^+ = -3.9$. 

The notional virtual origins for all three cases are included in table~\ref{tab:opposition}, along with the virtual origin for turbulence, predicted from their values using (\ref{eq:lTp}). The shift $\Delta U^+$ that these virtual origins would produce is also given in the table as the difference $\ell_U^+ - \ell_{T,pred}^+$ and compared to the shift $\Delta U^+$ measured from figure~\ref{fig:opposition}. {In the case of $v$ control and $v$-$w$ control,} the predicted $\Delta U^+$ agrees well with the measured one, but this is less so in the case of $w$ control. The discrepancy between the predicted and measured $\Delta U^+$, particularly in the case of $w$ control, could be caused by the control not successfully establishing a virtual origin for $w$ exactly halfway between the domain boundary and the detection plane. In fact, it appears to do so at some height below $y_d^+/2$. Note that in the present virtual-origin framework, the depth of the virtual origin perceived by $w$, i.e.\ $\ell_w^+$,  is set a priori, assuming that the shape of its r.m.s. velocity profile remained smooth-wall-like. However, the resulting apparent origin for $w$, as measured a posteriori, is not necessarily at $y^+ = - \ell_w^+$. Instead, as we argue in \S\,\ref{subsec:originForTurb}, the virtual origin perceived by the whole turbulence dynamics, and thus $w$, would be $y^+ = - \ell_T^+$, as can be appreciated for instance in  figure~\ref{fig:comp_uwv}(\textit{e}). As such, $\ell_w^+$ cannot be measured a posteriori from the r.m.s. profiles of the resulting flow, as only $\ell_T^+$ can. Nevertheless, from the resulting value of $\Delta U^+$ in the case of $w$ control, and the idea that the virtual origin perceived by $w$ appears to be the most limiting in terms of setting the virtual origin for turbulence, we deduce that the results are consistent with applying a priori $\ell_w^+ \approx -3$, instead of the notional value of $-3.9$. In addition, it can be observed from figure~\ref{fig:opposition}(\textit{b}) that the profile of $v^{\prime+}$ is also modified indirectly by the control of $w$ near the wall, suggesting that $\ell_v^+ \neq 0$, even though $v$ is not controlled directly. This highlights one of the key differences between the slip-length and opposition-control simulations. In the slip-length simulations, the virtual flow does not have to satisfy the incompressible Navier--Stokes equations below the plane in which the slip-length boundary conditions are applied, $y^+=0$. When the virtual origins are imposed, we simply assume that the r.m.s.\ velocity profiles extend below $y^+=0$ in a smooth-wall-like fashion but independently for each velocity component. This is in contrast to opposition control, where the flow must still satisfy continuity and the Navier--Stokes equations from the domain boundary up to the height of any virtual origin perceived by the flow, e.g.\ for $0\leq y^+ \leq -\ell_w^+$ for $w$ control. The underlying coupling between the three velocity components, and thus between their virtual origins, makes it difficult for our Robin-based framework to establish their locations a priori when a given velocity component is controlled, and therefore it is not always possible to predict accurately the origin for turbulence a priori. This area grants further research, but in any event it is worth noting that the underlying physical mechanism at play appears to be the same in both the opposition-control and the slip-length simulations. That is, each velocity component perceives a different apparent virtual origin, and this reduces further to a virtual origin perceived by the mean flow and a virtual origin perceived by turbulence. Then, if the virtual origin for the mean flow is deeper than the virtual origin for turbulence, the shift in the mean velocity profile $\Delta U^+$ is simply given by the height difference between the two, from (\ref{eq:DUlT}), and the turbulence above remains otherwise smooth-wall-like. {This also illustrates that the virtual origin for turbulence, even if determined by $v$ and $w$, may not always follow (\ref{eq:lTp}), but only when the effect of the control reduces to three different apparent velocity origins {that can be imposed through Robin boundary conditions}.}

\section{An eddy-viscosity model}\label{sec:eddyVisc}

Here, we present a simple model that captures the dependence of $\Delta U ^+$ on $\ell_U^+$ and $\ell_T^+$ (\ref{eq:DUlT}). For the smooth wall, we can approximate the turbulent mean velocity profile $U_{sm}^+(y^+)$ using an eddy-viscosity model for the Reynolds shear stress $-\overline{u'v'}^+ = (\nu_T/\nu)\mathrm{d}U_{sm}^+/\mathrm{d}y^+$ \citep[e.g.][]{vanDriest1956}, where $\nu_T(y^+)$ is the eddy viscosity representing turbulence. We will use $\nu_T^+(y^+)$ to refer to the normalised eddy viscosity $\nu_T(y^+)/\nu$. For channel flow, the total shear stress is linear:
\begin{equation}\label{eq:chan_bal}
\frac{\mathrm{d} U_{sm}^+}{\mathrm{d} y^+} -\overline{u'v'}^+= 
(1 + \nu_T^+)\frac{\mathrm{d} U_{sm}^+}{\mathrm{d} y^+} =
1 - \frac{y^+}{\Rey_\tau}.
\end{equation}
Two possible models for $\nu_T^+$ are those of \citet{vanDriest1956} and Cess~\citep[cf.][]{Reynolds1967}. The key difference between the two is that van Driest's model does not include the contribution from the wake, and so is only valid when $y/\delta \ll 1$. In the present study, since the focus is on the near-wall region of the flow, we choose to use van Driest's model for its relative simplicity. Noting that $y^+/\Rey_\tau = y/\delta$, the total stress in (\ref{eq:chan_bal}) becomes nearly uniform in the near-wall region, so we can write~\citep{vanDriest1956}:
\begin{equation}\label{eq:dUdy_eddy_visc}
\frac{\mathrm{d}U_{sm}^+}{\mathrm{d}y^+}  \approx f(y^+) = \frac{1}{1+\nu_T^+(y^+)},
\end{equation}
with
\begin{equation}\label{eq:eddy_visc_near_wall}
\nu_T^+(y^+)
  = \frac{1}{2} \left\lbrace 1+4\kappa^2 y^{+2}\left[1-\exp\left(-\frac{R(y^+)}{A}\right)\right]^2 \right\rbrace ^{1/2}
  -\frac{1}{2}.
\end{equation}
Here, $\kappa \approx 0.426$ and $A \approx 25.4$ \citep[cf.][]{Reynolds1967,delAlamo2006}, and $R(\cdot) \equiv \max(\cdot, 0)$ is the ramp function to ensure that the damping factor in the square brackets remains between 0 and 1 (in practice, the ramp function is regularised with $R(\cdot) = \log[1+\exp(\cdot)]$). The damping coefficient $A$ sets the thickness of the laminar sublayer by damping the contribution from turbulence just above the smooth wall, and thus also sets the log-law intercept $B$. For the above values of $\kappa$ and $A$, $B \approx 5.24$. For reference, we can check that (\ref{eq:dUdy_eddy_visc}), with this definition of $\nu_T/\nu$ (\ref{eq:eddy_visc_near_wall}), approaches $\mathrm{d} U^+/\mathrm{d}y^+ \sim 1/(\kappa y^+)$ for $y^+\gg1$ . In the limit of small $y/\delta$, (\ref{eq:eddy_visc_near_wall}) and the ensuing analysis also apply to other flows such as boundary layers. We can obtain the smooth-wall velocity profile by integrating (\ref{eq:dUdy_eddy_visc}), with the definition of $\nu_T^+$ given by (\ref{eq:eddy_visc_near_wall}), such that
\begin{equation}\label{eq:smooth_wall_int}
U_{sm}^+(y^+) = \int_0^{y^+} f(\xi^+) \,\mathrm{d}\xi^+,
\end{equation}
where $\xi^+$ is just the integration variable and $y^+$ is measured from the smooth wall where we have assumed $U_{sm}^+(0) = 0$ (as we are in the frame fixed to the wall). If $\nu_T^+ = 0$ in (\ref{eq:smooth_wall_int}), there is no turbulence and the flow stays laminar, $U^+ = y^+$.

\begin{figure}
\centering    
\includegraphics{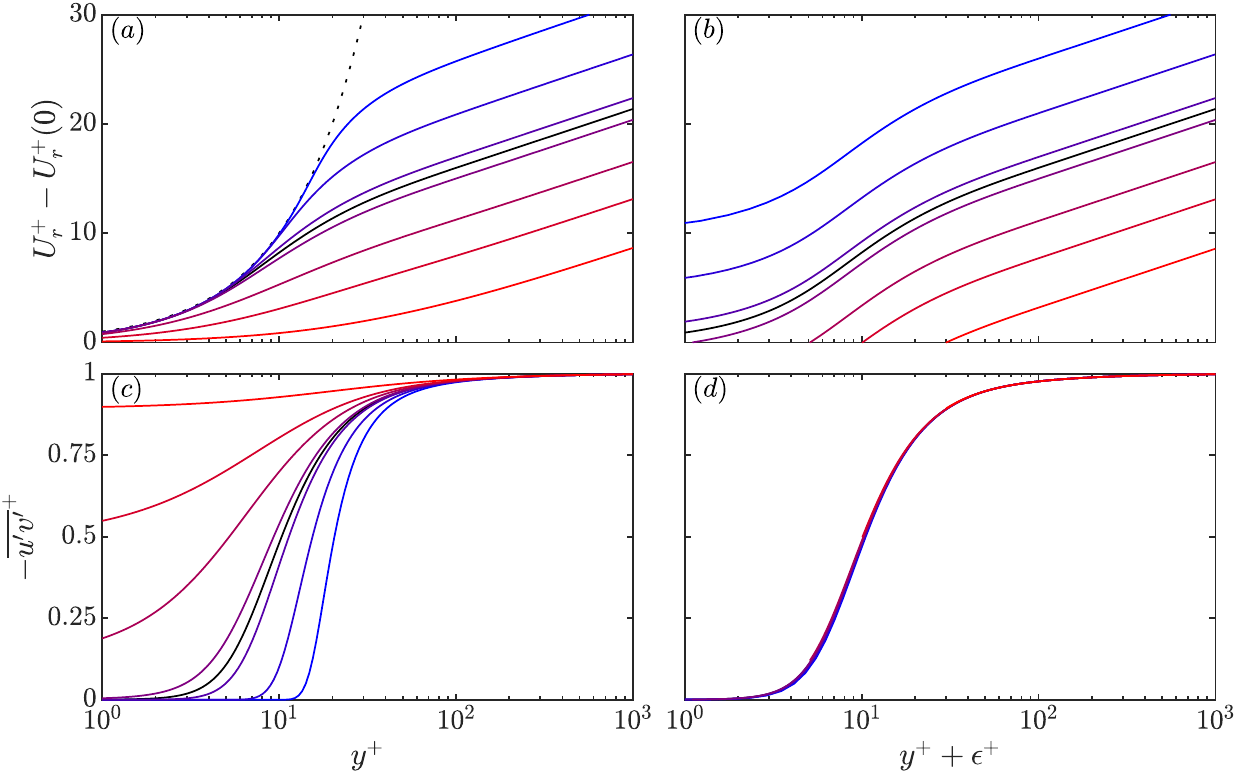}
\caption{Profiles of mean velocity $U_r^+$ from (\ref{eq:textured_wall_int}) (\textit{a},\textit{b}) and Reynolds shear stress $-\overline{u'v'}^+ = (\nu_T/\nu)\mathrm{d}U_r^+/\mathrm{d}y^+$ (\textit{c},\textit{d}), plotted relative to the origin for the mean streamwise flow $y^+$ (\textit{a},\textit{c}) and relative to the origin for turbulence $y^+ + \epsilon^+$(\textit{b},\textit{d}). Solid black lines, reference smooth-wall profiles, i.e.\ $\epsilon^+ = 0$; blue to red lines, $\epsilon^+ = [-10, -5, -1, 1, 5, 10, 30]$; dotted black line in (\textit{a}), laminar mean velocity profile, $U^+ = y^+$, for which $\nu_T/\nu = 0$.}
\label{fig:vo_profiles}
\end{figure}

We now apply the idea that the effect of a certain surface texture is to bring turbulence, represented by the eddy viscosity, closer to or farther from the reference plane $y^+=0$. In this model, this is achieved by shifting the eddy viscosity $\nu_T^+$ by $\epsilon^+$, say, in (\ref{eq:eddy_visc_near_wall}) and integrating to obtain the velocity profile above the textured wall $U_r^+$:
\begin{equation}\label{eq:textured_wall_int}
U_r^+(y^+) = U_r^+(0) + \int_0^{y^+} f(\xi^+ + \epsilon^+) \,\mathrm{d}\xi^+.
\end{equation}
Comparing (\ref{eq:textured_wall_int}) and (\ref{eq:smooth_wall_int}), we observe that if $\epsilon^+ = 0$, turbulence is placed as if a smooth wall were located at $y^+ = 0$, and the only effect of the texture is the Galilean transformation $U_r^+(0)$. Figure \ref{fig:vo_profiles}(\textit{a}) shows the mean velocity profile, less the Galilean transformation $U_r^+(0)$, for several values of $\epsilon^+$. If $\epsilon^+ > 0$, turbulence is brought closer to the wall, because $\nu_T^+$ activates for lower $y^+$. Similarly, if $\epsilon^+ < 0$, turbulence is lifted from the wall, because $\nu_T^+$ activates for higher $y^+$, as shown by the profiles of Reynolds shear stresses in figure \ref{fig:vo_profiles}(\textit{c}). Comparing figures \ref{fig:vo_profiles}(\textit{a}) and (\textit{c}), we observe increased velocity (drag reduction) for lifted turbulence and decreased velocity (drag increase) for lowered turbulence.

\begin{figure}
\centering    
\includegraphics[width=\textwidth]{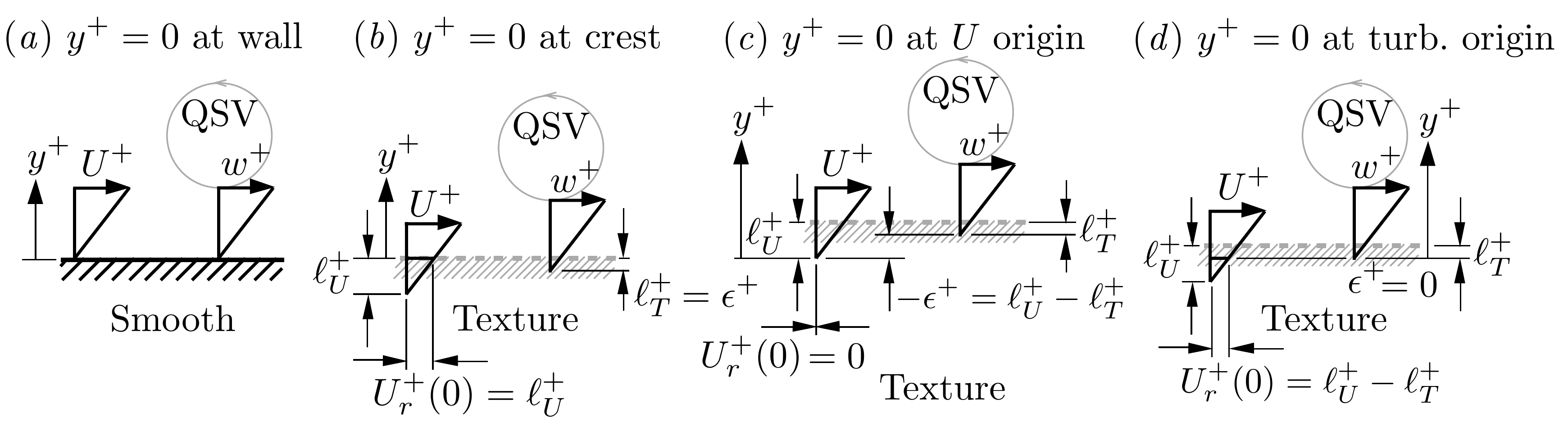}
\caption{Various choices for the reference plane $y^+=0$ when considering textured surfaces (\textit{b}, \textit{c}, \textit{d}) relative to the smooth wall (\textit{a}).These choices give rise to different streamwise slip velocities $U_r^+(y^+=0)$ and origins for turbulence $\epsilon^+$ relative to $y^+=0$, as indicated in the figures, and hence impact $\Delta U^+ \equiv U_r^+ - U_{sm}^+$ when evaluated at matched $y^+ \ll \infty$ using (\ref{eq:DUcess}). The depths $\ell_U^+$ and $\ell_T^+$ are not influenced by this choice; and their respective virtual origins are fixed relative to the texture.  QSV stands for quasi-streamwise vortices, used to represent the turbulence above walls. Note that $\partial u^+/\partial y^+ \approx 1$ near the smooth wall and surface texture.}
\label{fig:vo_ychoice}
\end{figure}

Unlike for the smooth wall, there are choices on where to locate $y^+ = 0$ for the textured wall. One convenient choice is at the crest of the textures. In this case, sketched in figure \ref{fig:vo_ychoice}(\textit{b}), $U_r^+(0)$ is the slip velocity evaluated at the crest and is equal to the height difference between the crest and the virtual origin for the mean flow, i.e.\ $U_r^+(0) = \ell_U^+$, and $\epsilon^+$ is the height difference between the crest and the virtual origin for turbulence, i.e.\ $\epsilon^+ = \ell_T^+$. To obtain the shift in mean velocity $\Delta U^+$, we subtract (\ref{eq:smooth_wall_int}) from (\ref{eq:textured_wall_int}) at matched $y^+$ (and choice of $y^+=0$):
\begin{equation}\label{eq:DUcess}
\Delta U^+(y^+) = U_r^+(y^+) - U_{sm}^+(y^+) = U_r^+(0)
  + \int_0^{y^+} \left\{f(\xi^+ + \epsilon^+)
   - f(\xi^+)\right\}\mathrm{d}\xi^+.
\end{equation}
If we instead chose $y^+=0$ to be the origin of the streamwise flow, as sketched in figure \ref{fig:vo_ychoice}(\textit{c}), then $U_r^+(0) = 0$ by definition and $\epsilon^+ = \ell_T^+-\ell_U^+$. Yet another choice for $y^+=0$ is the origin for turbulence, as sketched in figure \ref{fig:vo_ychoice}(\textit{d}), wherein $U_r^+(0) = \ell_U^+-\ell_T^+$ and $\epsilon^+ = 0$. This last choice is interesting because the integral in (\ref{eq:DUcess}) vanishes, and we obtain immediately $\Delta U^+ = \ell_U^+ - \ell_T^+$ for all $y^+$. We can see this in figure \ref{fig:vo_profiles}(\textit{b}), which portrays the mean velocity against the distance to the origin for turbulence $y^+ + \epsilon^+$, regardless of the choice of $y^+=0$. All profiles are parallel down into the viscous region, and so $\Delta U^+$ must be a constant for all $y^+$. Another point of consistency is that the modelled Reynolds shear stresses collapse when represented against the distance to the origin for turbulence $y^+ + \epsilon^+$, as shown in figure~\ref{fig:vo_profiles}(\textit{d}).

It is well known that in the log layer, where $y^+ \gg 1$, $\Delta U^+$ is independent of the choice for $y^+ = 0$. To see conditions under which this occurs in the present model, we can set $y^+ \rightarrow \infty$ (log layer) in the upper limit of integration in (\ref{eq:DUcess}) and find that the integral reduces to $-\int_0^{\epsilon^+}1/[1+(\nu_T/\nu)(\xi^+)]\,\mathrm{d}\xi^+ =-U^+(\epsilon^+)$, where $U$ is the smooth-wall velocity profile, cf.\ (\ref{eq:smooth_wall_int}), a somewhat surprising result. We know that the mean velocity profile for the smooth wall is $U^+(y^+) \sim y^+$ for $y^+\lesssim 5$, and so $U^+(\epsilon^+) \sim \epsilon^+$ for $-\infty<\epsilon^+ \lesssim 5$, assuming that the profile extends linearly below $y^+=0$. Physically, the lower limit on $\epsilon^+$ represents the idea that lifting turbulence away from the reference plane $y^+=0$, i.e.\ $\epsilon^+<0$, will allow the mean velocity profile to grow linearly, with unit gradient in wall units, up to $y^+ = -\epsilon^+$, regardless of the magnitude of $\epsilon^+$. That is, in (\ref{eq:textured_wall_int}) the integrand $ f(\xi^+ + \epsilon^+)$ will be unity when $y^+\leq-\epsilon^+$, since $\nu_T^+= 0$ for $y^+\leq 0$, as defined by (\ref{eq:eddy_visc_near_wall}). Substituting these results into (\ref{eq:DUcess}), we obtain $\Delta U^+ \sim U_r^+(0) - U^+(\epsilon^+) \sim U_r^+(0) - \epsilon^+$ for $y^+ \gg 1$ and $-\infty<\epsilon^+ \lesssim 5$.  This, in turn, reduces to $\Delta U^+ \sim \ell_U^+ - \ell_T^+$  for $y^+ = 0$ at crest, $\Delta U^+ \sim 0 - (\ell_T^+ - \ell_U^+)$ for $y^+ = 0$ at  the $U$-origin, and $\Delta U^+ \sim (\ell_U^+ - \ell_T^+) - 0$ for $y^+ = 0$ at the origin for turbulence. In other words, we observe that if the integral in (\ref{eq:DUcess}) is taken to $y^+ \rightarrow \infty$, $\Delta U^+ = \ell_U^+ - \ell_T^+$ is independent of the choice of reference plane $y^+ = 0$. Notably, \citet{Luchini1991} also demonstrated that, in the case of riblets, any `physically significant' measure of the effect of the texture on the flow should be independent of the choice of origin. However, in practice the log layer is not thick enough due to finite $\Rey_\tau$, and the integral in (\ref{eq:DUcess}) cannot be taken to infinity. In that case, equation (\ref{eq:DUcess})  indicates some sensitivity of $\Delta U^+$ to the choice of reference plane.
 
 \begin{figure}
\centering    
\includegraphics{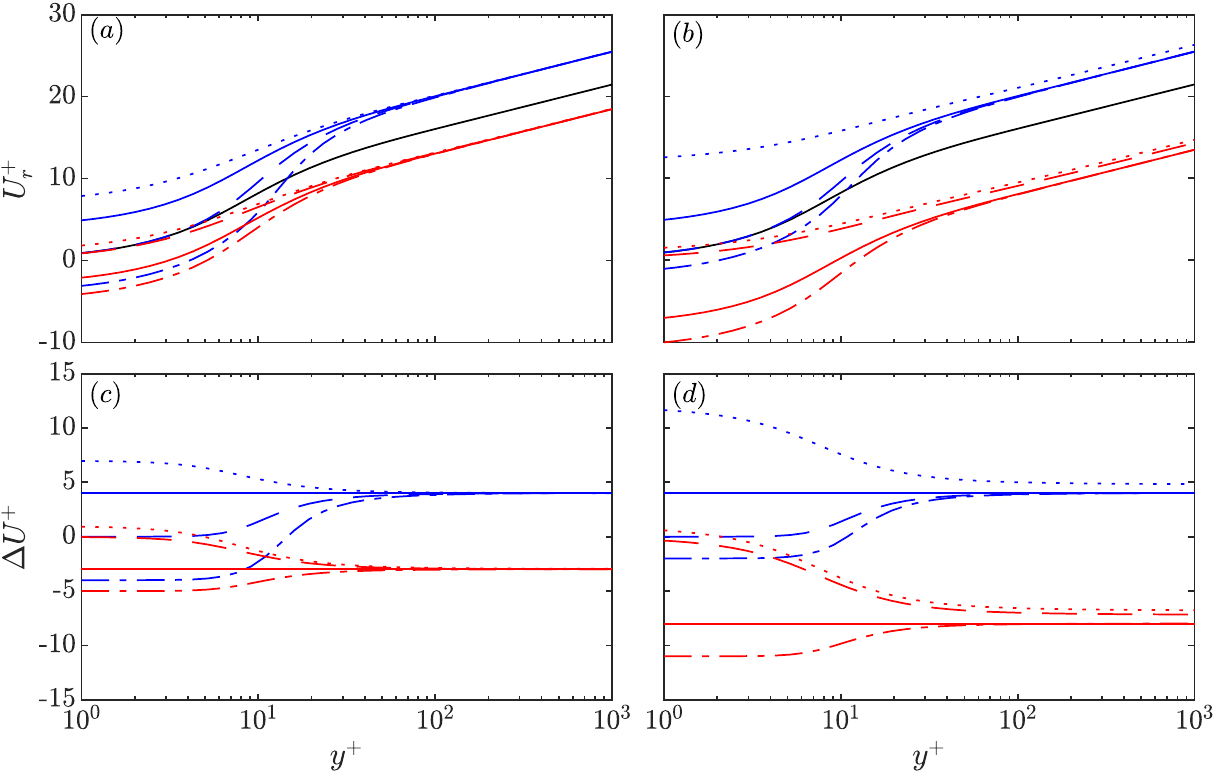}
\caption{Mean velocity profiles $U_r^+$ from (\ref{eq:textured_wall_int}) (\textit{a},\textit{b}) and variation of $\Delta U^+$ with $y^+$ from (\ref{eq:DUcess}) (\textit{c},\textit{d}) for different textured walls and various choices of $y^+ = 0$. In the left hand panels (\textit{a},\textit{c}): blue lines, $h^+ = 11$, with $\ell_U^+ = 7$ and $\ell_T^+ = 3$; red lines,  $h^+ = 6$, with $\ell_U^+ = 1$ and $\ell_T^+ = 4$. In the right-hand panels (\textit{b},\textit{d}): blue lines, $h^+ = 14$, with $\ell_U^+ = 12$ and $\ell_T^+ = 8$; red lines, $h^+ = 12$, with  $\ell_U^+ = 1$ and $\ell_T^+ = 9$. In all panels: dotted lines, $y^+ = 0$ at crest; dashed lines, $y^+ = 0$ at $U$-origin; solid lines, $y^+ = 0$ at origin for turbulence; dash-dotted lines, $y^+ = 0$ at valleys. The solid black line in (\textit{a},\textit{b}) denotes the reference smooth-wall profile.}
\label{fig:DU_cess_combined}
\end{figure}

We investigate this further by considering how  the value of $\Delta U^+$ is affected by the choice of $y^+=0$ and the upper limit of the integration in (\ref{eq:DUcess}), i.e.\ the height at which $\Delta U^+$ is measured. {We now also} include a fourth choice of $y^+=0$, where $y^+$ is measured from the notional valleys of the texture elements. In this reference frame, defining $h^+$ as the height of the elements in wall units, we would have $U_r^+(0) = \ell_U^+ - h^+$ and $\epsilon^+ = \ell_T^+ - h^+$, so that $\Delta U^+ \sim U_r^+(0) - \epsilon^+ = \ell_U^+ - \ell_T^+$ for $y^+ \rightarrow \infty$, as expected. In figure~\ref{fig:DU_cess_combined}(\textit{a}), profiles of $U_r^+$ from (\ref{eq:textured_wall_int}) are given for two hypothetical surfaces, one drag-reducing and one drag-increasing, with various choices of $y^+=0$. We see that, in each case, the profiles collapse only for large $y^+$ (in the log layer), as expected, and only then are they all parallel to the smooth-wall reference case. The profiles that have the origin for turbulence as $y^+=0$ are parallel to the smooth-wall profile for all $y^+$, as discussed in \S\,\ref{subsec:originForTurb}.  This is confirmed in figure~\ref{fig:DU_cess_combined}(\textit{c}), which shows the values of $\Delta U^+$ as a function of $y^+$ from (\ref{eq:DUcess}) for both hypothetical surfaces, for the various choices of $y^+=0$. The figure shows that $\Delta U^+$ is constant for all $y^+$ when the origin for turbulence is taken as $y^+=0$, whereas the curves for the other choices of $y^+=0$ asymptote to the `true' value only when $y^+\gtrsim 100$, i.e.\ in the log layer. This is consistent with the above analysis, where we demonstrated that $\Delta U^+$ is independent of the choice of $y^+=0$ when measured in the log layer. However, it highlights the potential for error when measuring $\Delta U^+$ in experiments or simulations too close to the wall, which could be the only option at low $\Rey_\tau$. That is, to precisely measure $\Delta U^+$ irrespective of the choice of  the reference $y^+=0$, the flow should exhibit a sufficiently thick log layer in the first place. While an exact definition of what this would require in practice is beyond the scope of this paper, this implies that $\Rey_\tau$ should be of the order of 500 or more, assuming the log layer is defined loosely as $80\nu/u_\tau \lesssim y \lesssim 0.3 \delta$~\citep{Sillero2013}. In contrast, if the origin for turbulence is taken as $y^+=0$, then accurate measurements of $\Delta U^+$ can be taken at any height, and thus at far lower values of $\Rey_\tau$.

From the various choices of reference plane, and ensuing definitions of $\epsilon^+$ (e.g.\ $\epsilon^+ = \ell_T^+$, $\epsilon^+ =0$, $\epsilon^+ = \ell_T^+ - \ell_U^+$ and $\epsilon^+ = \ell_T^+ - h^+$), we can infer the values of $\ell_U^+$ and $\ell_T^+$ that would violate the condition $-\infty<\epsilon^+ \lesssim 5$.  There is no need to consider a restriction on $h^+$, because when $y^+=0$ is taken as the texture valleys, $\epsilon^+ = \ell_T^+ - h^+$ will always be negative, since the origin for turbulence will always be above the valleys~\citep{Luchini1995}. {We see} immediately that the restrictions on $\ell_U^+$ and $\ell_T^+$ would be $\ell_T^+\gtrsim 5$ and $\ell_T^+ - \ell_U^+ \gtrsim 5$. Note that the former limit is relevant both in the regime of drag increase or drag reduction, while the latter limit is relevant for drag increase only. Once the limits are exceeded, $\Delta U^+$ would no longer necessarily be independent of the choice of $y^+=0$. Interestingly, we also deduced in \S~\ref{subsec:breakdown} from our slip-length simulations that the virtual-origins framework would break down for  $\ell_T^+\gtrsim 5$. Figures~\ref{fig:DU_cess_combined}(\textit{b},\textit{d}) portray the mean velocity profiles, along with the  variation of $\Delta U^+$ with $y^+$, for two hypothetical textured surface whose values of $\ell_U^+$ and $\ell_T^+$ result in $\epsilon^+ > 5$, depending on the choice of $y^+=0$. One surface satisfies $\ell_T^+\gtrsim 5$ only, while the other satisfies both $\ell_T^+\gtrsim 5$ and $\ell_T^+ - \ell_U^+ \gtrsim 5$. These figures demonstrate the potential for error when measuring $\Delta U^+$, even within the log layer. If $\epsilon^+$ is too large when the texture crests or the origin for the mean flow are taken as $y^+=0$, as often done in the literature, it is not possible to measure $\Delta U^+$ accurately at any height, regardless the magnitude of $\Rey_\tau$. Therefore, to consistently measure $\Delta U^+$ accurately for any texture and Reynolds number, $y^+$ should be measured from the origin for turbulence. However, if the texture valleys are taken as $y^+=0$, $\Delta U^+$ may still be measured accurately within the log layer, provided the Reynolds number is large enough.



\section{Conclusions}\label{sec:conclusions}

We have analysed the effect on turbulence of imposing different apparent virtual origins on the three velocity components, as some small-textured surfaces do. Examples of such surfaces are passive flow-control technologies, such as riblets, superhydrophobic surfaces or anisotropic permeable substrates. Our results show that, as long as the imposed virtual origins remain relatively small compared to the characteristic length scales of the near-wall turbulence cycle, the shift in the mean velocity profile, $\Delta U^+$, is determined by the offset between the virtual origin experienced by the mean flow and the virtual origin experienced by the turbulence, verifying equation (\ref{eq:DUlT}). The friction velocity that provides the scaling for the flow would not necessarily be the one derived directly from the surface drag, but from the total stress at the virtual origin for turbulence, $y=-\ell_T$. In practice, however, the difference between the two is negligible. In cases where the imposed virtual origins are no deeper than approximately 5 wall units, the turbulence remains essentially smooth-wall-like, other than for a wall-normal shift by $\ell_T^+$. We argue that it is possible to predict the virtual origin for turbulence a priori, and that lies between the virtual origins for the spanwise and wall-normal velocities, as expressed by equation (\ref{eq:lTp}). {The equation shows is that the only relevant parameters for determining the origin for turbulence are the relative positions of the virtual origins of $u$ and $w$ relative to the plane where $v$ appears to vanish. This is an extension to the original theory proposed by~\citet{Luchini1991} where, rather than on the difference between the virtual origins perceived by the tangential velocities, $\Delta U^+$ depends on their positions relative to that perceived by the wall-normal velocity, regardless of the plane taken as reference.} The virtual origin perceived by the streamwise velocity is essentially set by the streamwise slip length, $y=-\ell_x$. We have set this independently for the mean flow and the fluctuations, verifying that the one affecting $\Delta U^+$ is the origin for the mean flow. The virtual origin perceived by the streamwise velocity fluctuations, which are a proxy for the near-wall streaks, appears to be essentially inactive in setting the origin for turbulence, and hence has a negligible effect on the drag, at least in the regime where the origin perceived by the streaks is deeper than the origin perceived by the turbulence. In the opposite regime, the region occupied by the streaks eventually becomes too confined, and the near-wall turbulence no longer remains smooth-wall-like. Within the limits set by the above restrictions, it is possible to predict the shift in the mean velocity profile for a given textured surface using equations (\ref{eq:DUlT}) and (\ref{eq:lTp}). This analysis is valid for surfaces of small texture size, which do not alter the canonical nature of the turbulence, and we show that this result can also be predicted by introducing a virtual origin for turbulence into an a priori, eddy-viscosity model for the Reynolds shear stress. We also present exploratory results that suggest that the effect on the flow of opposition control, an active flow-control technique, can also be interpreted in terms of virtual origins.
\\

J.I.I.\ was supported by the Engineering and Physical Sciences Research Council (EPSRC) under a Doctoral Training Account, grant number EP/M506485/1. G.G.-d.-S.\ was supported by EPSRC grant EP/S013083/1. D.C.\ gratefully acknowledges the support of the Australian Research Council Discovery Project DP170102595. This work was also partly supported by the European Research Council through the $2^\text{nd}$ Coturb Madrid Summer Workshop. Computational resources were provided by the `Cambridge Service for Data Driven Discovery' operated by the University of Cambridge Research Computing Service and funded by EPSRC Tier-2 grant EP/P020259/1, {and by the DECI resource Cartesius, based in the Netherlands at SURFSara, with support from PRACE.}\\

Declaration of Interests. The authors report no conflict of interest.

\bibliographystyle{jfm}
\bibliography{jfmReferences}

\end{document}